\documentclass[a4paper,12pt, epsfig]{article}
\usepackage{epsfig}
\usepackage{amssymb,latexsym}
\usepackage{amsfonts}
\pdfoutput=1
\usepackage{amsmath}
\usepackage[colorlinks, linkcolor=blue, citecolor=blue, urlcolor=blue]{hyperref}
\usepackage{epstopdf}
\usepackage{graphicx}
\newskip\humongous \humongous=0pt plus 1000pt minus 1000pt

\newif\ifdtup

\catcode`\@=11

\@addtoreset{equation}{section}
\def\theequation{\arabic{section}.\arabic{equation}}

\def\@normalsize{\@setsize\normalsize{15pt}\xiipt\@xiipt
\abovedisplayskip 14pt plus3pt minus3pt%
\belowdisplayskip \abovedisplayskip
\abovedisplayshortskip \z@ plus3pt%
\belowdisplayshortskip 7pt plus3.5pt minus0pt}

\def\small{\@setsize\small{13.6pt}\xipt\@xipt
\abovedisplayskip 13pt plus3pt minus3pt%
\belowdisplayskip \abovedisplayskip
\abovedisplayshortskip \z@ plus3pt%
\belowdisplayshortskip 7pt plus3.5pt minus0pt
\def\@listi{\parsep 4.5pt plus 2pt minus 1pt
      \itemsep \parsep
      \topsep 9pt plus 3pt minus 3pt}}

\relax

\catcode`@=12

\catcode`\@=11

\def\section{\@startsection{section}{1}{\z@}{3.5ex plus 1ex minus
    .2ex}{2.3ex plus .2ex}{\large\bf}}

\def\thesection{\arabic{section}}
\def\thesubsection{\arabic{section}.\arabic{subsection}}

\def\appendix{\setcounter{section}{0}
  \def\thesection{Appendix \Alph{section}}
  \def\thesubsection{\Alph{section}.\arabic{subsection}}
  \def\theequation{\Alph{section}.\arabic{equation}}}
\def\SymBoxes#1#2#3#4{\newdimen\un@t \un@t#3%
\raisebox{#1}{\rule{#2\un@t}{#4}\hskip-#2\un@t% lower horizontal
\@tempdimb\un@t \advance\@tempdimb by-#4\@tempcntb#2\relax%
\@whilenum{\@tempcntb>0}\do{%                         % #2 vertical lines
\rule{#4}{\un@t}\hskip\@tempdimb \advance\@tempcntb by\m@ne}%
\hskip-#2\un@t \rule[\un@t]{#2\un@t}{#4}%
\rule[\un@t]{#4}{#4}\hskip-#4%             % upper horizontal line
\rule{#4}{\un@t}}\hskip-#4}                % rightest vertical line
\begin{document}
%\begin{letter}{~}

%%%%%%Define some new commands and  macros

\newcommand{\dd}{\textrm{d}}

\newcommand{\beq}{\begin{equation}}
\newcommand{\eeq}{\end{equation}}
\newcommand{\bea}{\begin{eqnarray}}
\newcommand{\eea}{\end{eqnarray}}
\newcommand{\beas}{\begin{eqnarray*}}
\newcommand{\eeas}{\end{eqnarray*}}
\newcommand{\defi}{\stackrel{\rm def}{=}}
\newcommand{\non}{\nonumber}
\newcommand{\bquo}{\begin{quote}}
\newcommand{\enqu}{\end{quote}}
\newcommand{\tc}[1]{\textcolor{blue}{#1}}
%%%%%%%%%%%%%%%%
\renewcommand{\(}{\begin{equation}}
\renewcommand{\)}{\end{equation}}
%%%%%%%%%%%%%%%%%%%%%%%%%%%%%%%%%% definitions
\def\de{\partial}
\def\Om{\ensuremath{\Omega}}
\def\Tr{ \hbox{\rm Tr}}
\def\rc{ \hbox{$r_{\rm c}$}}
\def\H{ \hbox{\rm H}}
\def\HE{ \hbox{$\rm H^{even}$}}
\def\HO{ \hbox{$\rm H^{odd}$}}
\def\HEO{ \hbox{$\rm H^{even/odd}$}}
\def\HOE{ \hbox{$\rm H^{odd/even}$}}
\def\HHO{ \hbox{$\rm H_H^{odd}$}}
\def\HHEO{ \hbox{$\rm H_H^{even/odd}$}}
\def\HHOE{ \hbox{$\rm H_H^{odd/even}$}}
\def\K{ \hbox{\rm K}}
\def\Im{ \hbox{\rm Im}}
\def\Ker{ \hbox{\rm Ker}}
\def\const{\hbox {\rm const.}}
\def\o{\over}
\def\im{\hbox{\rm Im}}
\def\re{\hbox{\rm Re}}
\def\bra{\langle}\def\ket{\rangle}
\def\Arg{\hbox {\rm Arg}}
\def\exo{\hbox {\rm exp}}
\def\diag{\hbox{\rm diag}}
\def\longvert{{\rule[-2mm]{0.1mm}{7mm}}\,}
\def\a{\alpha}
\def\b{\beta}
\def\e{\epsilon}
\def\l{\lambda}
\def\ol{{\overline{\lambda}}}
\def\ochi{{\overline{\chi}}}
\def\th{\theta}
\def\s{\sigma}
\def\oth{\overline{\theta}}
\def\ad{{\dot{\alpha}}}
\def\bd{{\dot{\beta}}}
\def\oD{\overline{D}}
\def\opsi{\overline{\psi}}
\def\dag{{}^{\dagger}}
\def\tq{{\widetilde q}}
\def\L{{\mathcal{L}}}
\def\p{{}^{\prime}}
\def\W{W}
\def\N{{\cal N}}
\def\hsp{,\hspace{.7cm}}
\def\hspp{,\hspace{.5cm}}
\def\bo{\ensuremath{\hat{b}_1}}
\def\bfo{\ensuremath{\hat{b}_4}}
\def\co{\ensuremath{\hat{c}_1}}
\def\cfo{\ensuremath{\hat{c}_4}}
\def\th#1#2{\ensuremath{\theta_{#1#2}}}
\def\c#1#2{\hbox{\rm cos}(\th#1#2)}
\def\s#1#2{\hbox{\rm sin}(\th#1#2)}
\def\cp#1#2#3{\hbox{\rm cos}^#1(\th#2#3)}
\def\sp#1#2#3{\hbox{\rm sin}^#1(\th#2#3)}
\def\ctp#1#2#3{\hbox{\rm cot}^#1(\th#2#3)}
\def\cpp#1#2#3#4{\hbox{\rm cos}^#1(#2\th#3#4)}
\def\spp#1#2#3#4{\hbox{\rm sin}^#1(#2\th#3#4)}
\def\t#1#2{\hbox{\rm tan}(\th#1#2)}
\def\tp#1#2#3{\hbox{\rm tan}^#1(\th#2#3)}
\def\m#1#2{\ensuremath{\Delta M_{#1#2}^2}}
\def\mn#1#2{\ensuremath{|\Delta M_{#1#2}^2}|}
\def\u#1#2{\ensuremath{{}^{2#1#2}\mathrm{U}}}
\def\pu#1#2{\ensuremath{{}^{2#1#2}\mathrm{Pu}}}
\def\meff{\ensuremath{\Delta M^2_{\rm{eff}}}}
\def\an{\ensuremath{\alpha_n}}
\newcommand{\Z}{\ensuremath{\mathbb Z}}
\newcommand{\R}{\ensuremath{\mathbb R}}
\newcommand{\rp}{\ensuremath{\mathbb {RP}}}
\newcommand{\vac}{\ensuremath{|0\rangle}}
\newcommand{\vact}{\ensuremath{|00\rangle}                    }
\newcommand{\oc}{\ensuremath{\overline{c}}}
\renewcommand{\cos}{\textrm{cos}}
\renewcommand{\sec}{\textrm{sec}}
\renewcommand{\sin}{\textrm{sin}}
\renewcommand{\cot}{\textrm{cot}}
\renewcommand{\tan}{\textrm{tan}}
\renewcommand{\ln}{\textrm{ln}}

\newcommand{\Vol}{\textrm{Vol}}

\newcommand{\half}{\frac{1}{2}}

%%%%%%%%%%%%%%%%%%%%%%%Changed%%%%%%%%%%%%%%%%%%%%%%%%%%%%%
\def\changed#1{{\bf #1}}
%\def\changed#1{ #1}
%%%%%%%%%%%%%%%%%%%%%%%%%%%%%%%%%%%%%%%%%%%%%%%%%%%%%%%%%

\begin{titlepage}
%\begin{flushright}
%IFUP-TH/2011-??
%\end{flushright}
%\bigskip

\def\thefootnote{\fnsymbol{footnote}}

\begin{center}
{\large {\bf
Showering Cosmogenic Muons in A Large Liquid Scintillator
  } }
%\end{center}

\bigskip

\bigskip

{\large \noindent Marco Grassi$^{1,2}$\footnote{\texttt{mgrassi@ihep.ac.cn}},  Jarah
Evslin$^{3,4}$\footnote{\texttt{jarah@ihep.ac.cn}}, Emilio
Ciuffoli$^{3,5}$\footnote{ciuffoli@ihep.ac.cn} and Xinmin Zhang$^{3,5}$\footnote{\texttt{xmzhang@ihep.ac.cn}} }
\end{center}

\renewcommand{\thefootnote}{\arabic{footnote}}

\vskip.7cm

\begin{center}
\vspace{0em} {\em  1) Institute of High Energy Physics (IHEP), CAS, Beijing 100049, China\\
2) INFN - Sezione di Roma, P.le Aldo Moro 2, 00185 Rome, Italy\\
3) Theoretical Physics Center for Science Facilities, IHEP, CAS,
   %Chinese Academy of Sciences, 
Beijing 100049, China\\
4)  Institute of Modern Physics, CAS, NanChangLu 509, Lanzhou 730000, China\\
   5) Theoretical physics division, IHEP, CAS, %\\  %Chinese Academy of Sciences,
    % P.O. Box 918(4), 
Beijing 100049, China\\

 {}}

%\vspace{0em} {\em  { 1) TPCSF, IHEP, Chinese Acad. of Sciences\\
%2) Theoretical physics division, IHEP, Chinese Acad. of Sciences\\
%YuQuan Lu 19(B), Beijing 100049, China}}

\vskip .4cm

\vskip .4cm

\end{center}

\vspace{1.3cm}

\noindent
\begin{center} {\bf Abstract} \end{center}

\noindent
We present the results of FLUKA simulations of the propagation of cosmogenic muons in a 20 kton spherical liquid scintillator detector underneath 700 to 900 meters of rock.  A showering muon is one which deposits at least 3 GeV in the detector in addition to ionization energy.  We find that 20 percent of muons are showering and a further 10 percent of muon events are muon bundles, of which more than one muon enters the detector.  In this range the showering and bundle fractions are robust against changes in the depth and topography, thus the total shower and bundle rate for a given experiment can be obtained by combining our results with an estimate for the total muon flux.  One consequence is that a straightforward adaptation of the full detector showering muon cuts used by KamLAND to JUNO or RENO 50 would yield a nearly vanishing detector efficiency.

\vfill

\begin{flushleft}
{\today}
%\vspace{1cm}
\end{flushleft}
\end{titlepage}
%\bigskip

\hfill{}
%\bigskip

%\tableofcontents

\setcounter{footnote}{0}

\section{Introduction} \label{intro}
In the next decade, the experiments JUNO \cite{juno} and RENO 50 \cite{reno50} and perhaps LENA \cite{lena} will employ the largest liquid scintillator detectors ever constructed to detect antineutrinos from a variety of sources.  While these detectors will be underground to shield them from cosmogenic muons, the first two will be at the relatively modest depths of 700 and 900 meters respectively.  The showers induced by some of these muons will react with the carbon in the liquid scintillator and create ${}^9$Li and ${}^8$He some of whose decays yield the same double coincidence \cite{cr} used to identify antineutrinos via inverse beta decay (IBD).  In previous experiments, such as KamLAND, the resulting background was largely eliminated by vetoing events occurring soon after the passage of cosmogenic muons \cite{kamlandback}.  However in the case of these two new, large detectors such a veto is delicate as the time between muon events is comparable to both the lifetimes of ${}^9$Li and of ${}^8$He.  In this paper we present the results of a series of simulations of muons in such detectors.  The results of these simulations on the one hand indicate that KamLAND's cuts cannot be straightforwardly applied to JUNO and RENO 50 but on the other hand can be used to determine the efficiency of any new veto scheme.

We used the FLUKA simulation package \cite{fluka} to simulate the propagation of cosmogenic muons and antimuons inside of a 20 kton spherical detector consisting of a LAB-based liquid scintillator.  This corresponds to the detector proposed for the experiment JUNO, while the RENO 50 detector will be 18 kton and cylindrical.  We considered several different muon spectra, corresponding to various overburdens, topographies and cosmogenic muon distribution models.  We will report results obtained using the cosmogenic muon distribution of Ref.~\cite{bundle} which is illustrated in Fig.~\ref{initfig}.   As $\mu^-$ and $\mu^+$ interactions in the detector result in different isotope production rates \cite{lindley}, the ratio of $\mu^+$ to $\mu^-$ will affect our results.  We will assume an energy independent ratio of 1.37 $\mu^+$ per every $\mu^-$ corresponding to the value measured  by Kamiokande for 1.2 TeV muons in Ref.~\cite{antimu}.  In Sec.~\ref{docciasez} we will describe the distribution of the muon's energy deposition in the scintillator, subtracting the deposition due to ionization which does not contribute significantly to the ${}^9$Li and ${}^8$He yield.   In our next paper we will describe the effect of the ${}^9$Li and ${}^8$He yields on the physics goals of these experiments, which will allow an optimal choice of veto strategies for these experiments given certain assumptions regarding the yet unknown tracking abilities of the detectors.

The unprecedented size of these detectors leads to a second consequence.  As shown in Ref.~\cite{bundle}, interactions of cosmic rays with the atmosphere often yield muon bundles consisting of multiple muons which travel in nearly the same direction.  At the relevant depths the separations of muons are typically of order 10 meters.  This means that in the case of a relatively small detector like KamLAND typically only a single muon will be observed in each event.  As a result less than 5\% of the muon events at KamLAND resulted in the detection of multiple muons \cite{kamlandback}.  On the other hand  JUNO, RENO 50 and LENA are all much larger than 10 meters and so a majority of muon bundles will result in multiple muons impacting the detector.  This will present several challenges.  First, it will be nearly impossible to determine how much energy was deposited by which muon, and so to determine whether one or both muons is showering.  Next, it will be more difficult to track the individual muons, although this tracking is important because full detector cuts in such a large detector weigh heavily upon the detector efficiency.  Third, while we will find that the probability of an individual muon showering is about 20\%, the probability that at least one muon is showering in a two muon event is about 30\%.  In Sec.~\ref{bundlesez} we will estimate the muon bundle frequency in various cases of interest.  

Finally in Sec.~\ref{issez} we will determine the expected rates of the spallation isotopes ${}^9$Li and ${}^8$He.  We will see that these background rates are greater than the expected reactor neutrino signals, and so some veto scheme is necessary.

\section{Showering Muons} \label{docciasez}

We have used the initial muon energy distribution at several depths of interest as the input of a FLUKA simulation which determined the energy deposited by each muon in a spherical (inner) detector containing 20 kton of the liquid scintillator.    As the composition of the outer detector has not yet been fixed, we made the crude approximation that outside of the detector lies a vacuum and so did not consider showers beginning in this region.

Muons lose energy via various processes.  Of these, ionization occurs at an average rate per track length which is independent of muon's energy so long as this energy far exceeds the muon's mass, as it does for muons that traverse the entire detector.  On the other hand, other processes, such as bremsstrahlung, pair production and photonuclear interactions lead to energy dissipation rates which increase with increasing muon energy.  The charged particles produced or liberated in these interactions themselves lose energy via ionization, called secondary ionization.  When we write {\it{ionization}} we will always mean primary ionization, that created by the muon directly.

\begin{figure} %[!tph]
\begin{center}
\includegraphics[width=4.2in]{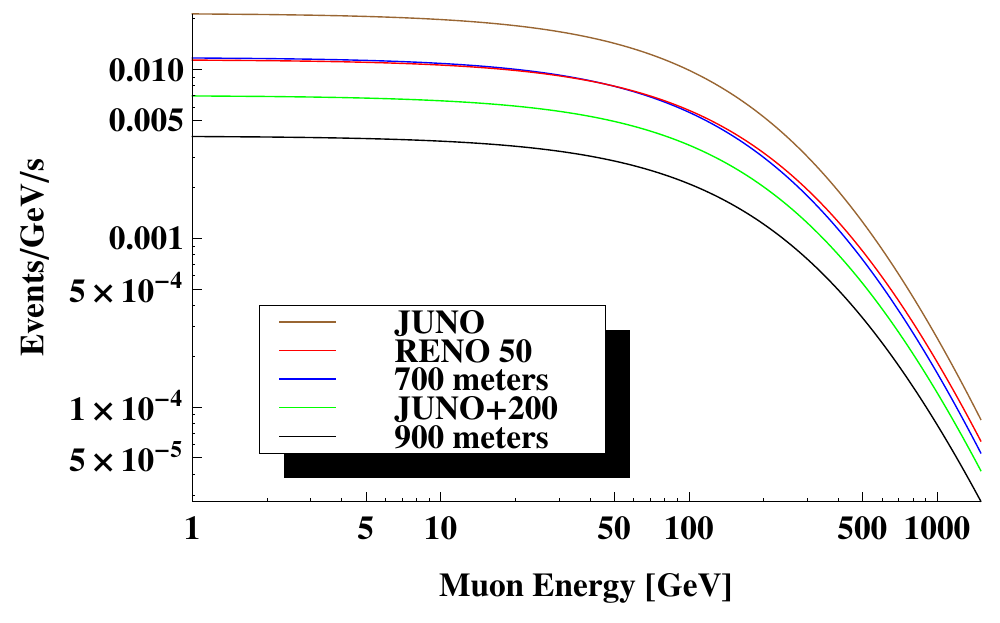}
\caption{The energy distribution of single cosmogenic muon events at a depth of 700 meters and 900 meters of rock assuming a flat surface and also at the proposed locations of the experiments JUNO and RENO 50, as well as 200 meters under the location proposed for JUNO}
\label{initfig}
\end{center}
\end{figure}

Most electrons liberated by ionization have a low energy and do not travel far from the track.  However, the energy distribution has a long high energy tail consisting of electrons, called $\delta$ rays, which can travel far from the track.  The former, due to their low energy, do not produce isotopes and so are irrelevant to our backgrounds.  Furthermore they are liberated at such a high rate that the amount of energy lost by the muon by this processes can be calculated quite precisely by simply multiplying the track length by a constant.  In the case of the scintillator relevant for JUNO and RENO 50, this constant is 1.43 MeV/cm.  Therefore, when we refer to the energy deposited minus ionization, we mean the energy deposited minus the track length multiplied by 1.43 MeV/cm.  This is a quantity available to the experimenter assuming that the muon is well tracked.  FLUKA separately calculates the $\delta$ ray production, where $\delta$ rays are defined as primary ionization electrons with energies greater than 100 keV.  Although technically these too are created by ionization, we will include them in our ionization subtracted plots.  In summary, when we state that we subtract the ionization energy we always mean that we subtract 1.43 MeV/cm which to a very good approximation is equal to the expected primary ionization energy not counting the high energy $\delta$ ray tail.  We will refer to the ionization subtracted energy as the showering energy.  

\begin{figure} %[!tph]
\begin{center}
\includegraphics[width=4.2in]{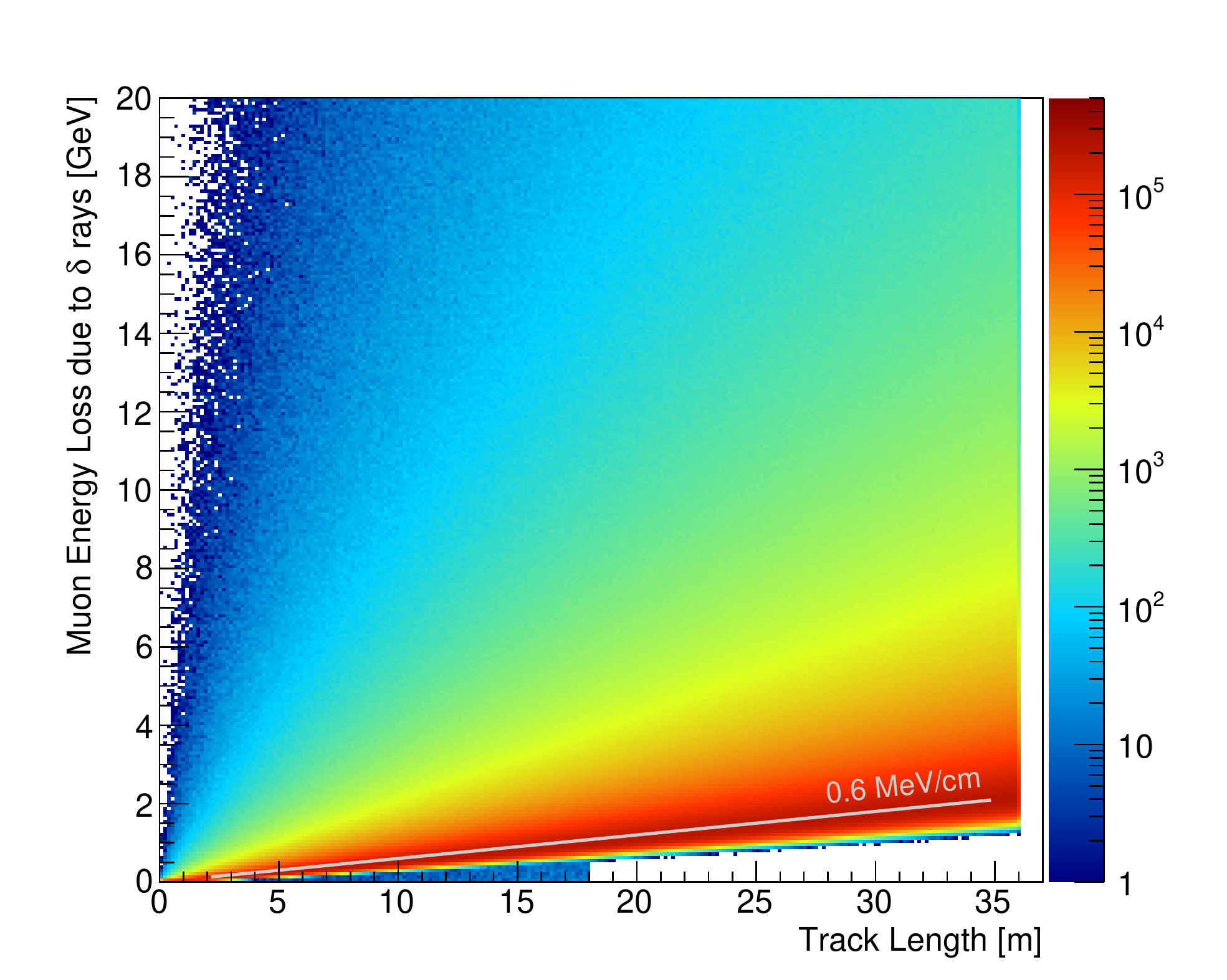}
\caption{The energy deposition of $\delta$ rays in the JUNO detector as a function of track length.  The average deposition is roughly 0.88 MeV/cm, while the most common deposition is only about 0.6 MeV/cm.}
\label{deltafig}
\end{center}
\end{figure}

In fact $\delta$ rays, except for perhaps the very far end of their high energy tail, also are not sufficiently energetic to lead to produce isotopes.  Furthermore in a large detector like JUNO, a single $\mu$ creates so many $\delta$ rays that the energy lost due to $\delta$ ray emission can be roughly approximated by a constant energy deposition per track length, although the scatter is nonnegligible as can be seen in Fig.~\ref{deltafig}.  Therefore one may expect the production of spallation isotopes to be correlated with the energy deposition with both the ionization and $\delta$ ray energies subtracted.  In Figs.~\ref{lispectra} and \ref{cumfig} we will therefore consider the total energy, the showering energy which is equal to the showering energy minus the ionization energy and also the showering minus minus the expected $\delta$ ray energy.  The average $\delta$ ray energy deposition is $0.88$ MeV/cm however this is dominated by a long high-energy tail which contains a relatively small portion of the events.  The $\delta$ ray energy deposition from the vast majority of muons is about $0.6$ MeV/cm, as is visible in Fig.~\ref{deltafig}.  Therefore in these figures we subtract not the average $\delta$ ray energy, which is highly sensitive to the long tail, but rather the peak deposition energy $0.6$ MeV/cm. 

%The energy deposited via ionization does not contribute to the creation of ${}^9$Li and ${}^8$He and also can be reliably subtracted if the track of the muon is known.  Therefore we will not consider the energy deposited by ionization.  It may be subtracted in two different ways, which yield essentially the same results.  First of all, the ionization energy is proportional to the path length of the muon and so, once the path is known, it can be calculated and subtracted using a constant of proportionally appropriate for the liquid scintillator.  This will be the only method which may be employed in the data analysis of the actual experiment.  Second, FLUKA automatically separated the energy deposition in categories, of which one is ionization, and so one may simply not include the ionization deposition in the energy loss extrapolated from FLUKA.  In the case of multimuon events, the detector cannot separate the energies deposited by the various muons, thus the multimuon event case is quite different from the single muon events. 

The diameter of the detector considered is about 35 meters, which is much larger than typical separation of muons in a muon bundle \cite{bundle}.  As a result if one muon from a bundle strikes the detector, we will see that it is likely that all muons in the bundle strike the detector and so muon bundles lead to multimuon events. 
This is quite different from the case of KamLAND, whose detector diameter is of order a bundle size and so in general at most one muon from each bundle strikes the detector.  As a result the single muon rate at KamLAND includes most bundle events.  Multimuon events will be the subject of Sec.~\ref{bundlesez}.  For simplicity, in this section we will consider only single muon events.

Following \cite{bundle}, the flux of incident muons arriving from a zenith angle $\theta$ is 
\beq
K(h,\theta)=7.2\times 10^{-3}h^{-1.927}\cos(\theta)e^{(-0.581h+0.034)/\cos(\theta)}  \label{keq}
\eeq
where $h$ is the weighted difference in altitude measured in kilometers between the detector and the surface.  The material will always be standard rock, and so we define a weighted altitude $h$ as 2.64 times the true altitude $h_r$.  While this note was written for a flat surface in which case $h$ is a constant, we will also consider detectors under mountains in which case $h$ may be a function of $\theta$.  This is the correct prescription for generalizing the results of \cite{bundle} to nontrivial topographies because muons arising from different angles $\theta$ are independent of one another, and so $h$ can be chosen independently for each value of $\theta$.  

As a result of the factor of $\cos(\theta)$, Eq.~(\ref{keq}) yields the flux of muons crossing a horizontal surface of unit area.  Our detector is not a horizontal surface, it is a sphere.  Therefore the correct angle in our case is not $\theta$ but rather the angle between the normal to the detector and the direction from which the muon arrives.  However we may simplify our expression for the flux by noting that, for a spherical detector, whatever the angle $\theta$ of the incoming muon the total cross sectional area of the detector is $\pi r^2$ where $r$ is the detector's radius.  We set $r$ to 18 meters although the inner detector radius is actually 17.7 meters.  As a result of spill in of the muon showers the fiducial volume for muon events is likely to be greater than 18 meters.  Thus the incident flux upon the detector from an angle $\theta$ is $\pi r^2 K(h,\theta)/\cos(\theta)$ where the factor of $\cos(\theta)$ has been removed as $\pi r^2$ is the area of the detector perpendicular to the velocity of the incident muon.  Integrating this quantity over the angles from which the muons arrive yields the single muon event rate
\beq
R=\pi r^2 \int d\Omega K(h,\theta)/\cos(\theta) = 2 \pi^2 r^2 \int_0^{\pi/2} d\theta K(h,\theta)\tan(\theta) \label{rate}
\eeq  
where we recall that, if the terrain is not flat, $h$ will be a nontrivial function of $\theta$.

The muon flux per unit energy at a given angle $\theta$ is
\bea
%\frac{dN(E,h,\theta)}{dE}&=&e^{0.42 h\sec({\theta})(-2.961+0.232\ln(h))}[E+\epsilon(1-e^{-0.42 h\sec({\theta})})]^{0.232\ln(h)-3.961}\nonumber\\
\frac{dN(E,h,\theta)}{dE}&=&[E+\epsilon(1-e^{-0.42 h\sec({\theta})})]^{0.232\ln(h)-3.961}\nonumber\\
\epsilon&=&0.0304e^{0.359h}\sec({\theta})-0.0077h+0.659 
\eea
up to a normalization term which can be fixed by demanding that the integral of $N$ over $E$ yield the incident flux $\pi r^2  K(h,\theta)/\cos(\theta)$.   $E$ is the energy in units of TeV.  Fixing this normalization and integrating over solid angles in the upper half of the detector's surface one finds that the muon rate per unit energy is
\bea
\frac{dR(E)}{dE}&=& 2 \pi^2 r^2 \int_0^{\pi/2} d\theta G(h,\theta) K(h,\theta) \frac{dN(E,h,\theta)}{dE}\tan(\theta) \label{enrate}\nonumber\\
G(h,\theta)&=&(2.961-0.232\ln(h))[\epsilon(1-e^{-0.42 h\sec({\theta})})]^{(2.961-0.232\ln(h))}
\eea  
where again we recall that for a nonflat topography $h$ is a function of $\theta$.   The rate per energy is reported Fig.~\ref{initfig} for detectors 700 meters and 900 meters underneath a flat standard rock surface, at the preferred JUNO and RENO 50 sites and also 200 meters below the preferred JUNO site.

\begin{figure} %[!tph]
\begin{center}
\includegraphics[width=4in]{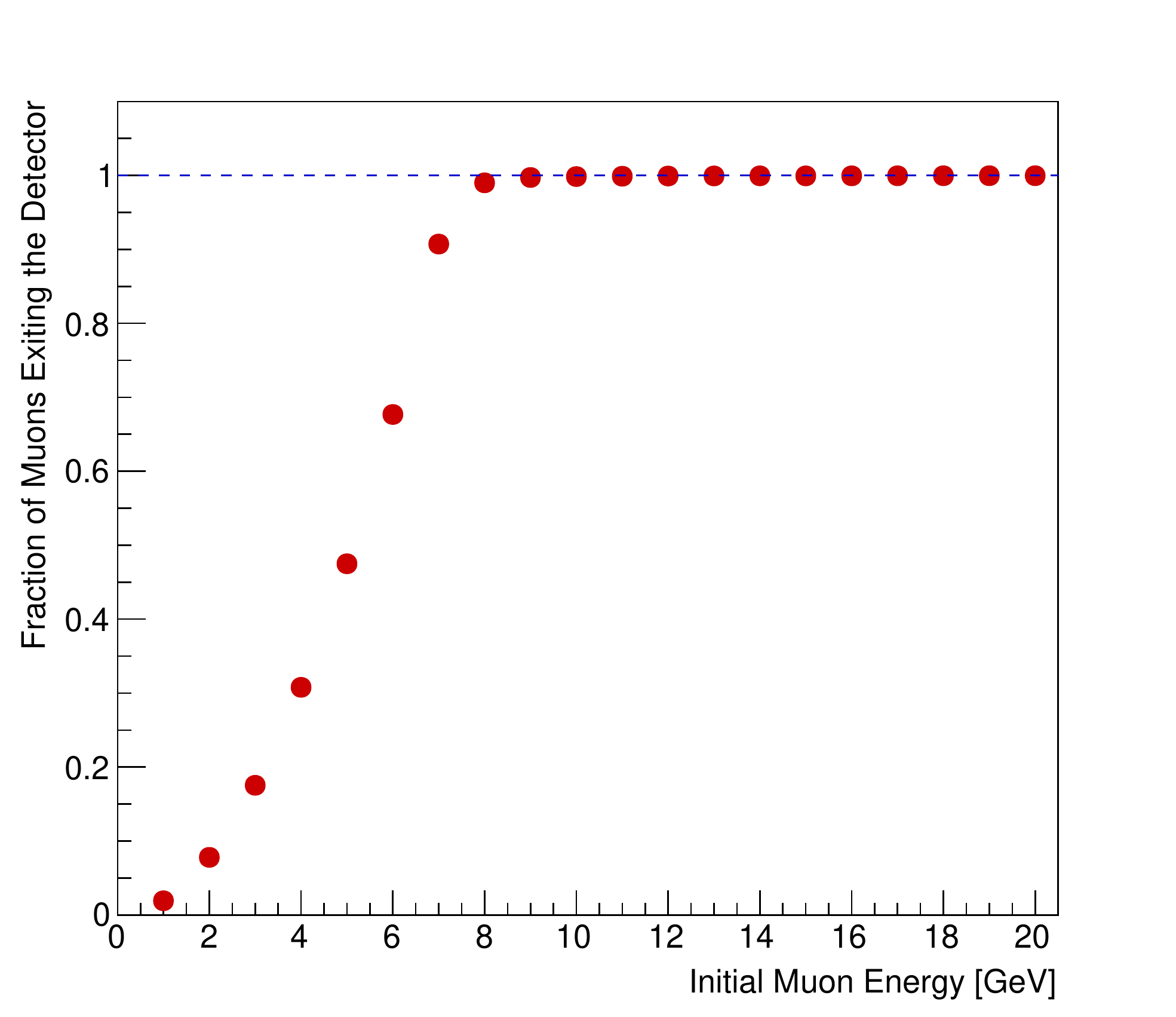}
\caption{The fraction of muons that traverse the detector as a function of muon energy.  When the energy is less than 10 GeV the constant energy loss per path length caused by primary ionization including $\delta$ ray emission can stop muons inside of the detector, while higher energy muons essentially never stop in the detector.}
\label{sopraviv}
\end{center}
\end{figure}

%The normalization term depends upon $h$ and $\theta$ and may be fixed numerically.  

%After normalizing $N$ it may be integrated over solid angles $d\Omega$ to arrive at the total flux as a function of energy, reported in 

Given the initial energy and the trajectory of a muon with respect to the detector, FLUKA yields the energy deposited via various channels.  We have simulated muons incident upon the detector with various impact parameters, weighted according to their likelihood such that all muons pass through the inner detector.   As can be seen in Fig. \ref{sopraviv}, unless the muon energy is less than 10 GeV essentially all muons that enter the detector also exit.

\begin{figure} %[!tph]
\begin{center}
\includegraphics[width=4in]{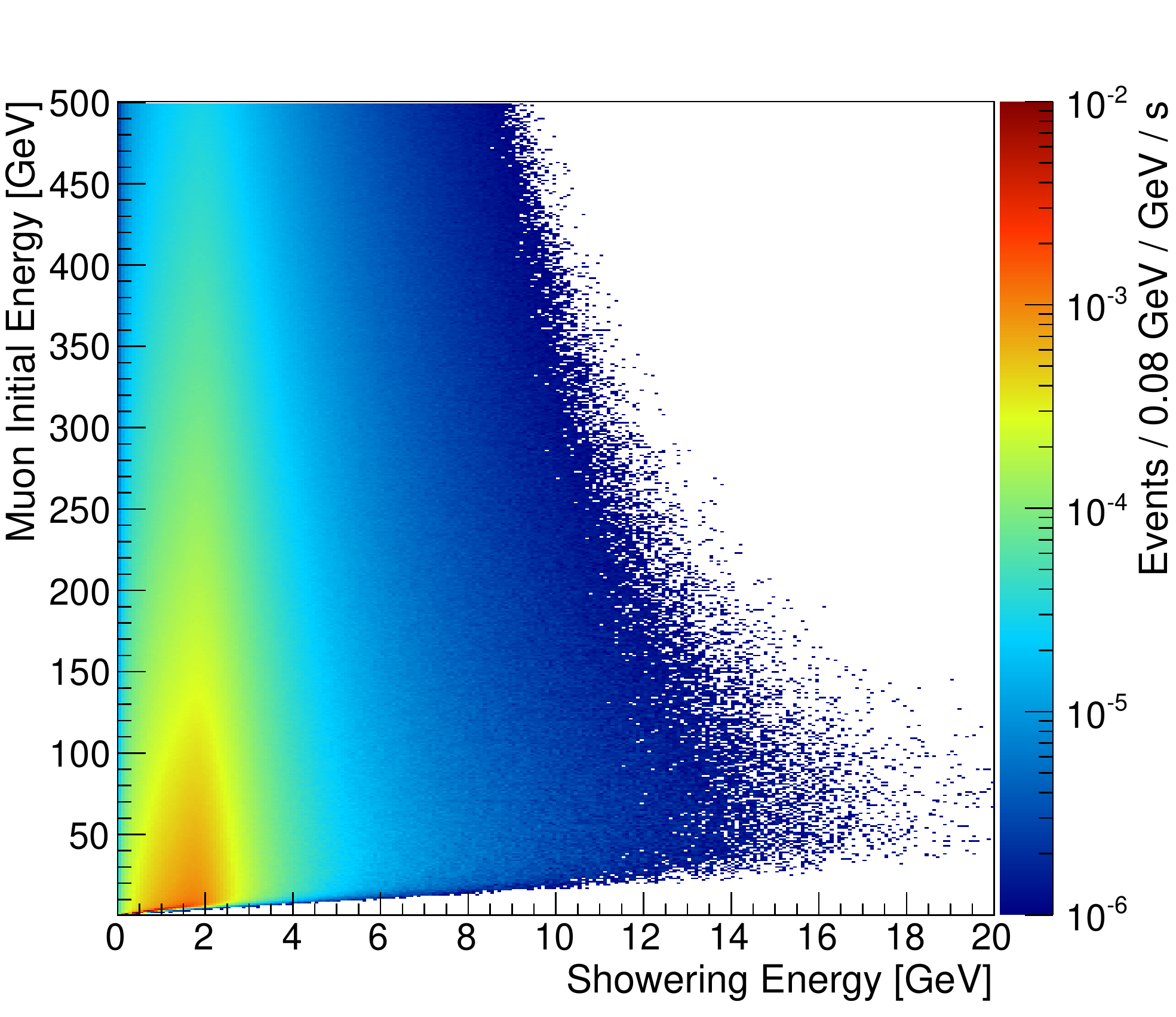}
\caption{The initial and deposited (not counting ionization) energy of cosmogenic muons at the preferred site for the JUNO experiment.}
\label{initfinal}
\end{center}
\end{figure}

\begin{figure} %[!tph]
\begin{center}
\includegraphics[width=3in]{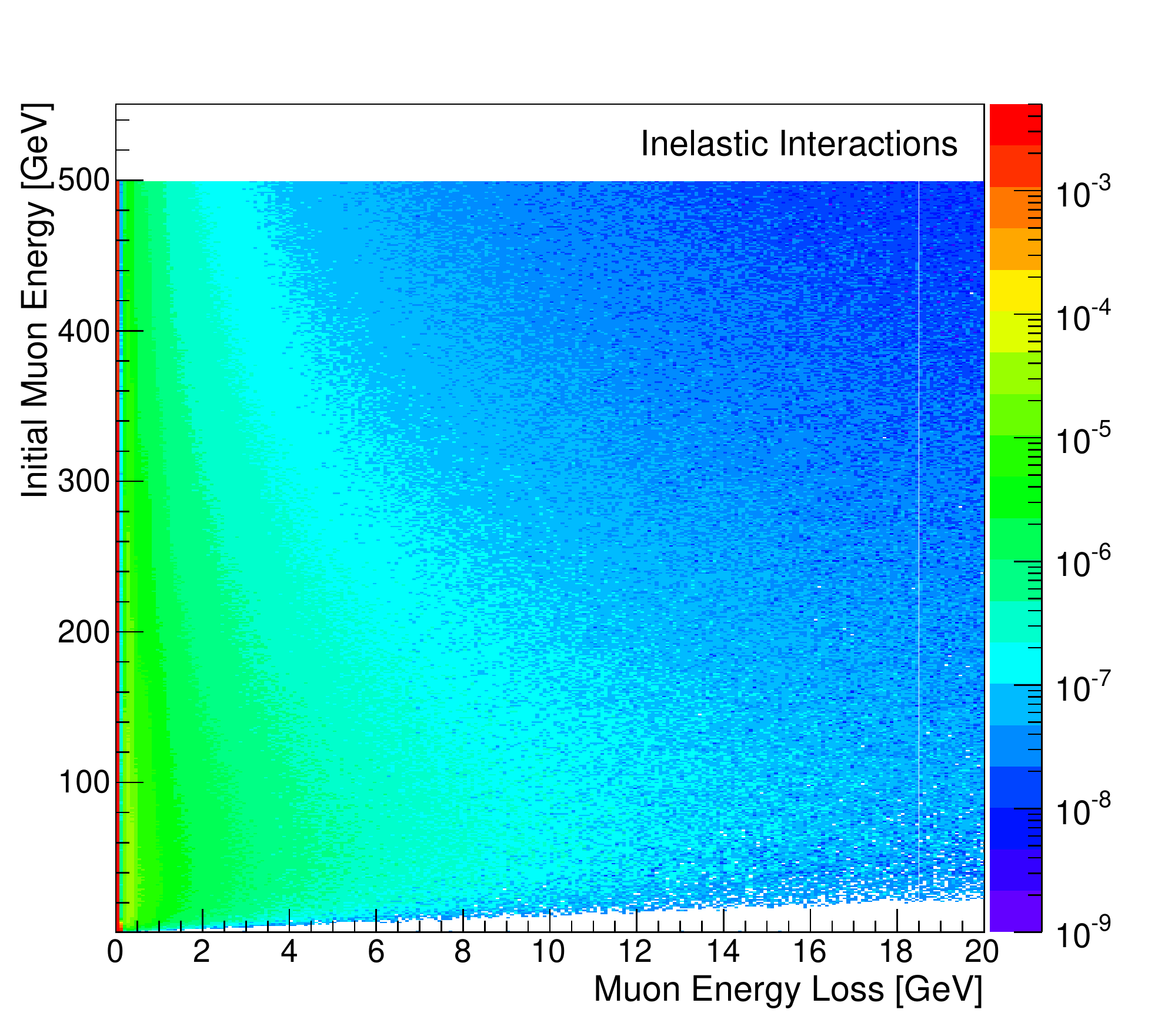}
\includegraphics[width=3in]{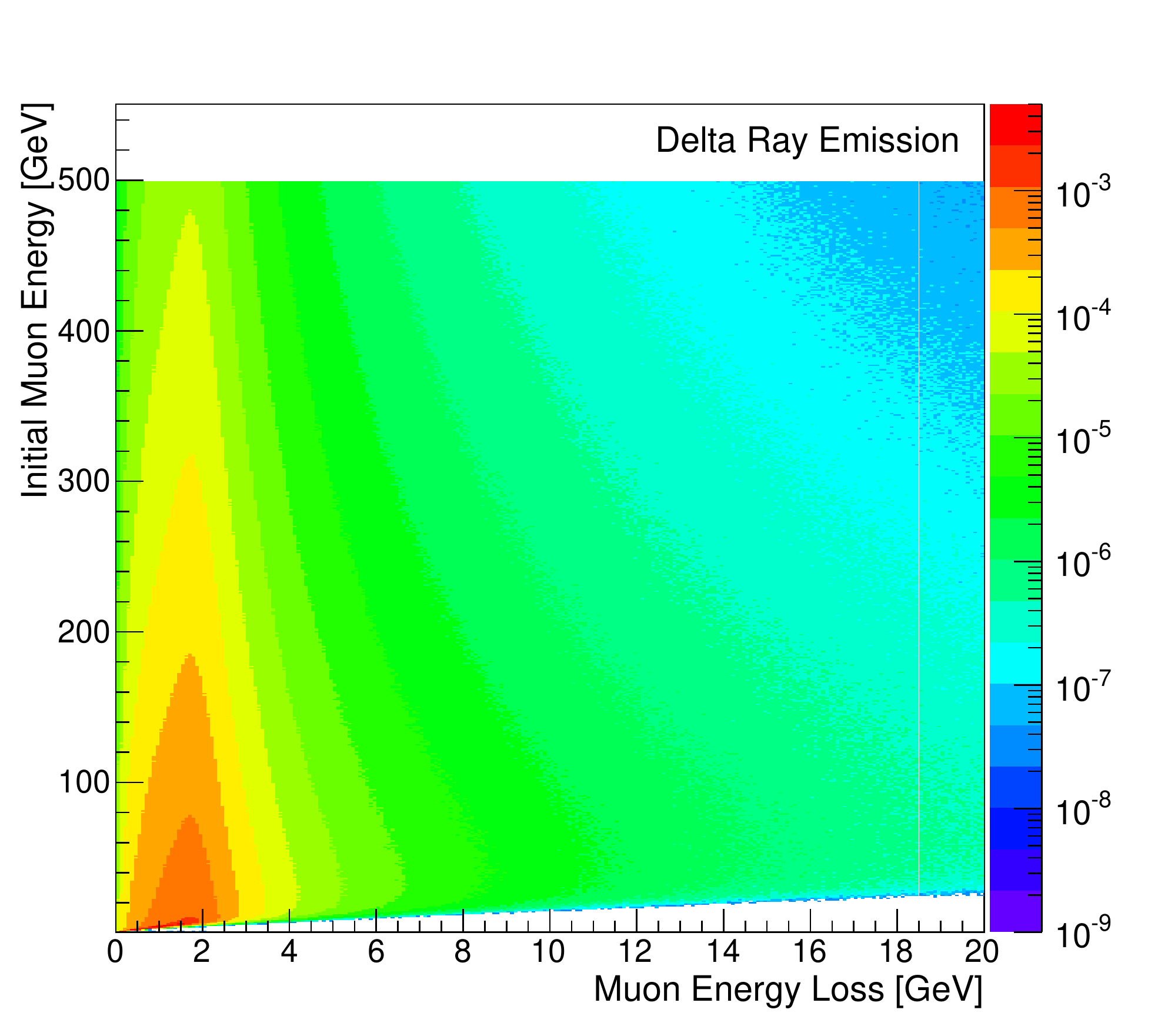}
\includegraphics[width=3in]{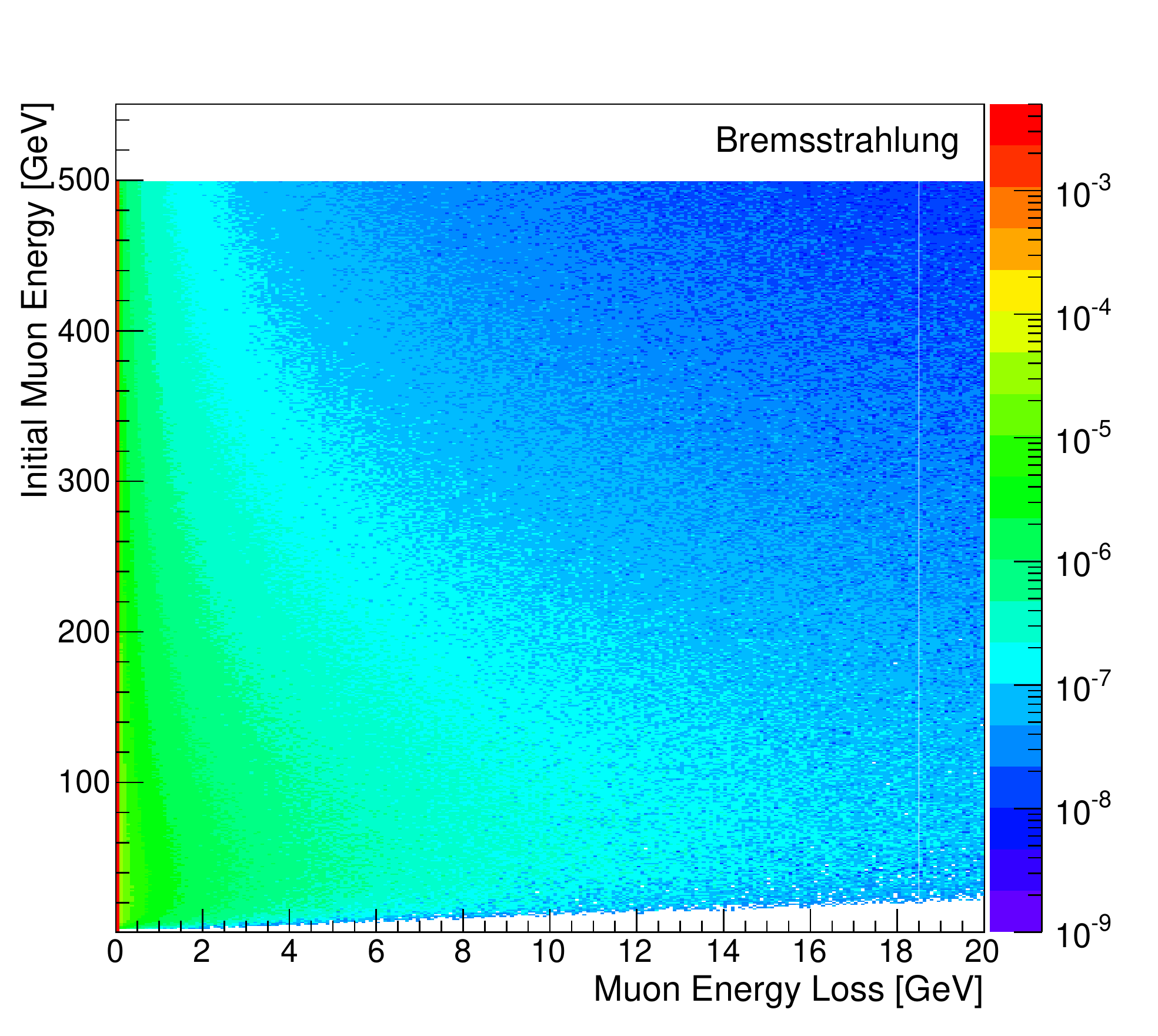}
\includegraphics[width=3in]{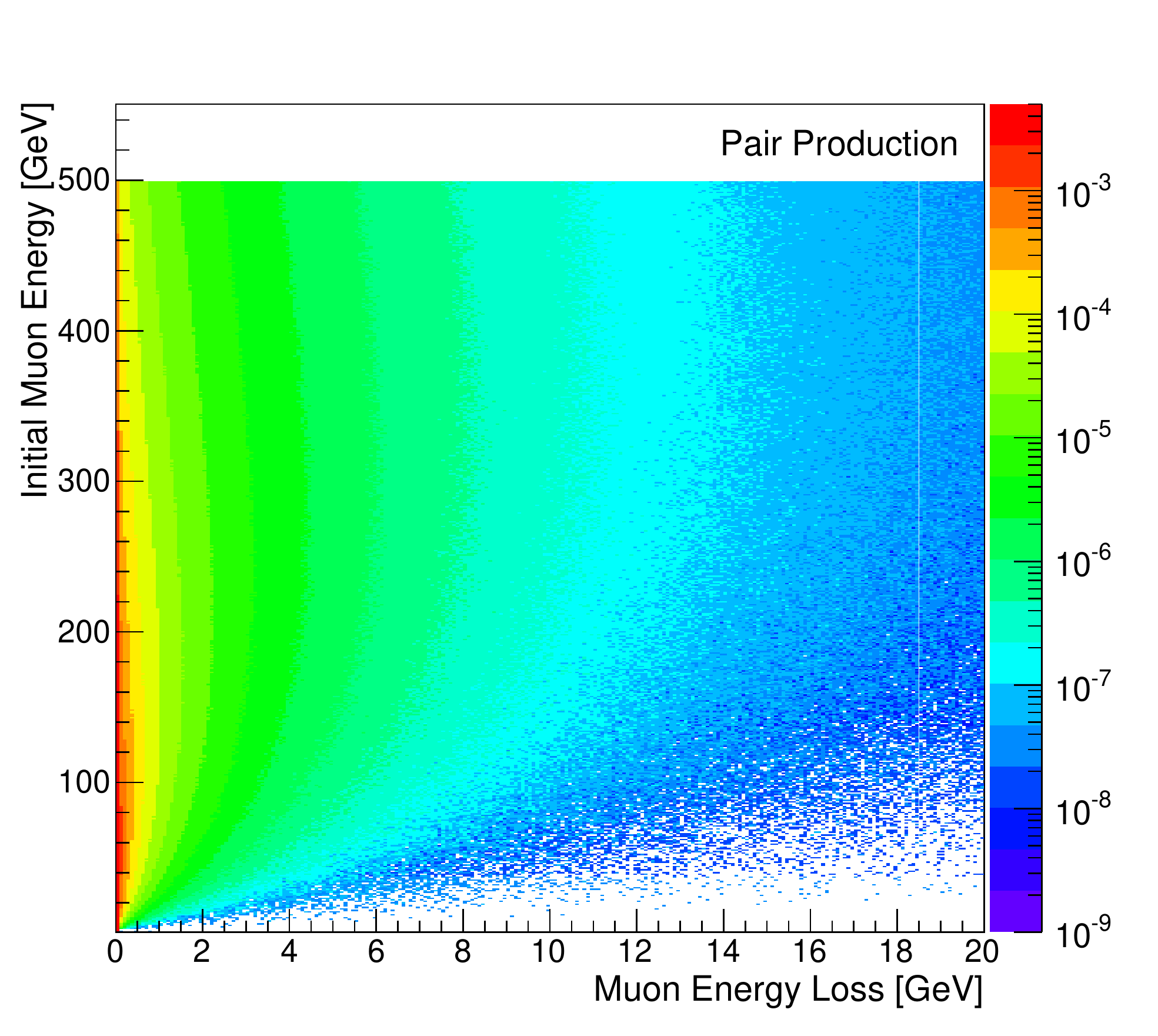}
\caption{As in Fig. \ref{initfinal} but now decomposed into various processes.  The main inelastic process consists of photonuclear interactions.  One can observe that $\delta$ ray emission is responsible for most energy deposition and in particular for the spectrum of the energy deposition.  These deposit a constant energy per path length of the track.  As seen in Fig. \ref{sopraviv}, when a muon's energy is less than about 10 GeV  it cannot cross the detector and so the energy deposited is proportional to the track length which is proportional to the muon energy.}
\label{initfinaldelta}
\end{center}
\end{figure}

\begin{figure} %[!tph]
\begin{center}
\includegraphics[width=4in]{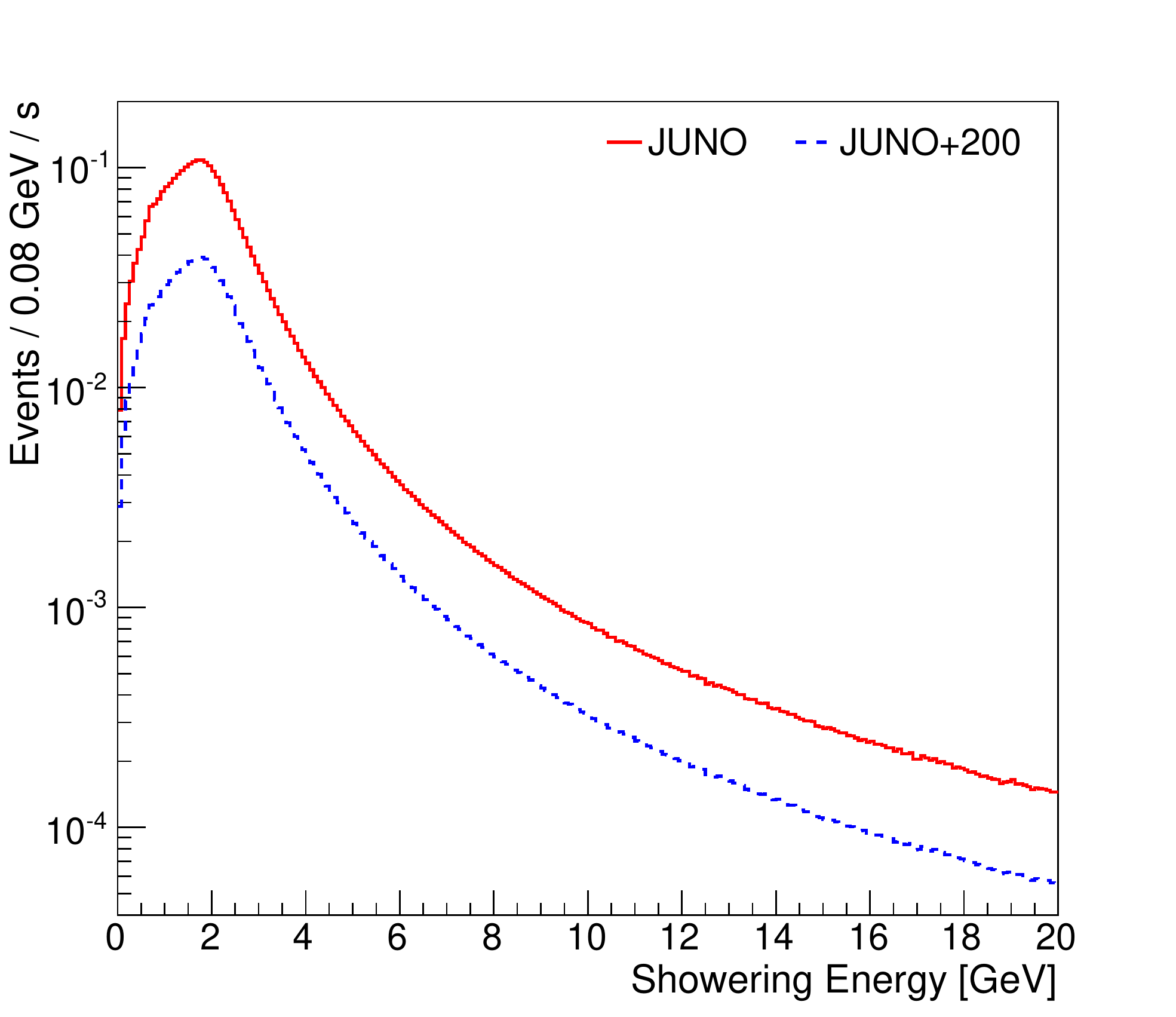}
\includegraphics[width=4in]{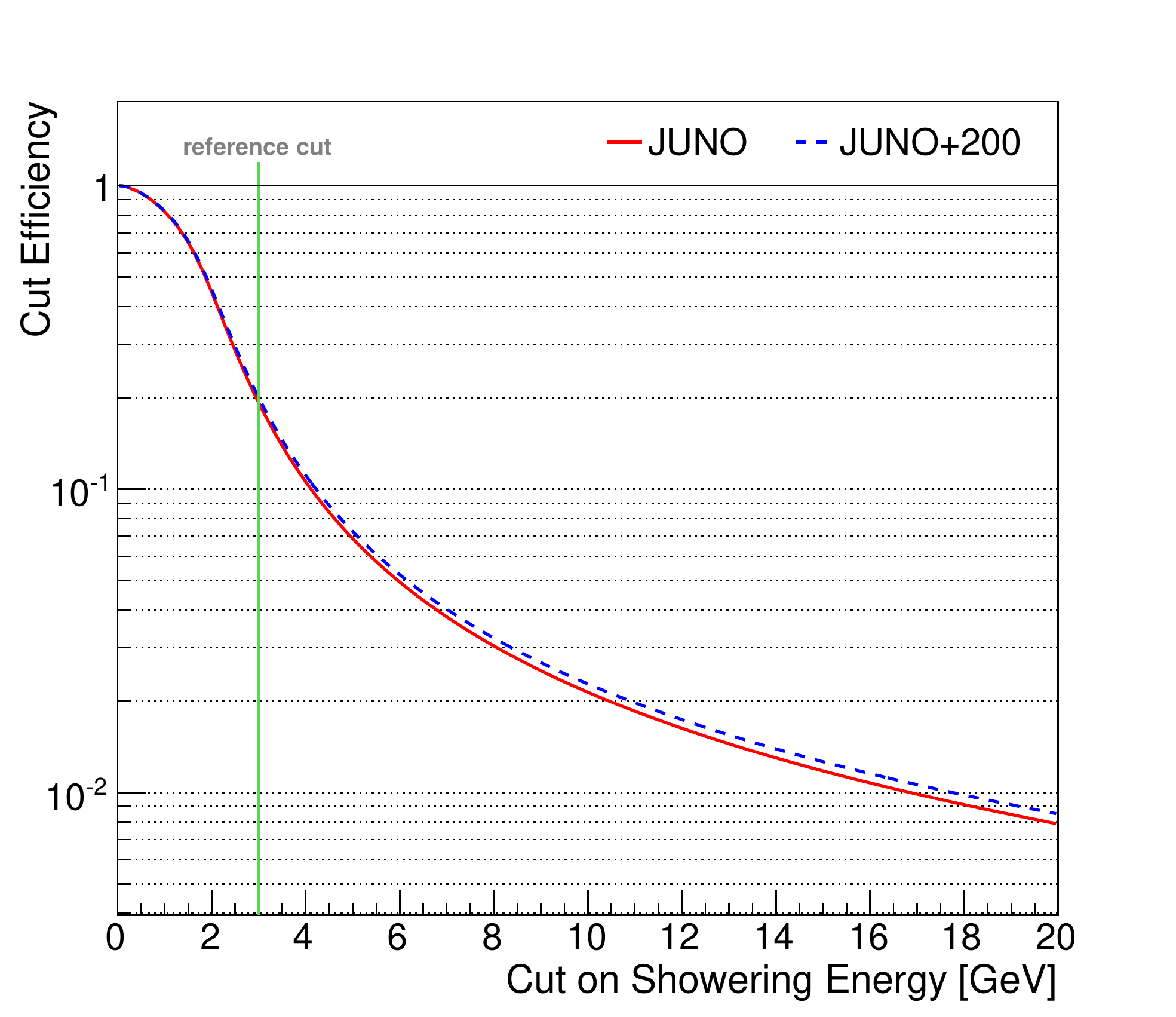}
\caption{Top panel: The rate of muons depositing a given amount of energy is shown, not including the energy deposited by ionization, at the preferred site for the JUNO experiment and 200 meters deeper.  Bottom panel: The normalized integral of the top panel, showing the fraction of muons which deposit more than a given threshold energy.  Note that only the high energy tail of this curve is sensitive to the overhead burden in this range.  The green line illustrates that 20 percent of muons deposit more than 3 GeV in addition to that deposited by ionization.}
\label{finalfig}
\end{center}
\end{figure}

Each datapoint in Fig.~\ref{initfinal} corresponds to the initial energy and the energy deposited not counting ionization of one of our simulated muons.  One may observe that when the muon energy is greater than 10 GeV, the most likely energy deposition is 2 GeV.   In Fig. \ref{initfinaldelta} the this deposition is decomposed into various processes and it can be seen that the 2 GeV maximum is due to the production of $\delta$ rays.  This is consistent with the observation that energy of this maximum is independent of the initial muon energy.   Had bremsstrahlung, pair production or photonuclear interactions instead dominated the deposition they would have led to a peak deposition energy which would have increased with the muon energy. Integrating this distribution over the initial energy one arrives at the upper panel of Fig.~\ref{finalfig}, which displays the distribution $\rho(E)$ of energies deposited by muons at the preferred site for the JUNO experiment and 200 meters deeper.   

Integrating $\rho(E)$ from $E$ to $\infty$ and normalizing the result to unity we obtain the lower panel of Fig.~\ref{finalfig}.  This displays the fraction of muons for which the deposited energy, not counting that deposited by ionization, is greater than each fixed level $E$.  The definition of a showering muon used by the KamLAND collaboration is one which deposits 3 GeV of energy in addition to that deposited by ionization.   As can be seen in Fig 7.18 of Ref.~\cite{dwyer}, this definition is useful as, in the case of a KamLAND sized detector, the vast majority of ${}^9$Li and ${}^8$He is produced by showering muons.  The corresponding results for JUNO are shown in Fig. \ref{lispectra}.  One readily sees that at JUNO and RENO 50 the contribution of nonshowering muons to ${}^9$Li and ${}^8$He production will have little effect on the science goals of the experiment even if they are not vetoed at all.   The lower panel of Fig.~\ref{finalfig} shows that the fractional energy deposition distribution of the muons is quite insensitive to the depth in the range considered and in fact the showering fraction is extremely robust.  For a 20 kton detector using KamLAND's definition of a showering muon, as had been anticipated in Ref.~\cite{menufact} about {\bf{20\% of muons are showering at these depths}}, 200 meter variations in the depth and indeed even 200 meter changes in the topography have little effect on this robustly determined fraction.    This is our main result.

\section{Muon Bundles} \label{bundlesez}

In this section, we will use the parametric formulae of Ref.~\cite{bundle} to determine the multimuon event rates at 20 kton spherical liquid scintillator detectors in various settings. We will then calculate the probability that more than a muon of a single bundle will hit the detector.  The $m$-muon flux is
\beq
\Phi(h,\theta,m)=\frac{K(h,\theta)}{m^{(-0.0771h^2+0.524h+2.068)e^{0.03e^{0.47h} \sec(\theta)}}}. \label{phieq}
\eeq
The $m$-muon rate can be calculated as in Eq.~(\ref{rate})
\beq
R(m) = 2 \pi^2 r^2 \int_0^{\pi/2} d\theta \Phi(h,\theta,m)\tan(\theta).
\eeq  
This is the muon bundle rate, which is the rate with which the axes of muon bundles strike the detector. To relate this number to an observable quantity, one can observe that $mR(m)$ is the rate at which muons which are part of $m$-muon bundles strike the detector.  Here $m$ is the total number of muons in the bundle, whether or not they all strike the detector.

Of course, one is interested not in the number of muons in a bundle but in the number of muons which actually strike the detector.  We will now calculate this in the case of 2-muon bundles, while in the case of bundles with 3 or more muons we will make the crude approximation that at least 2 always strike the detector.  Ref.~\cite{bundle} provides the probability density $f(R)=dN/dR$ of the separation $R$ between a muon in a bundle and the axis of the shower that generated the bundle 
\beq
f(R)=C\frac{R}{(R+R_0)^\alpha}
\eeq
where 
\bea
\alpha(h,m)&=& (-0.448m+4.969)e^{(0.0194m+0.276)h} \nonumber \\ 
 R_0(h,\theta,m)&=&\frac{\alpha(h,m)-3}{2}\frac{(-1.786 m +28.26)h^{-1.06m}}{e^{10.4(\theta-1.3)}+1}
 \nonumber\\
 C&=&(\alpha-1)(\alpha-2)R_0^{\alpha-2}.
\eea
Below we will need the probability density not for the distance from the axis, but rather for the distance $D$ between two muons
\beq
f'(D)=\frac{dN}{dD} .
\eeq
Consider a 2 muon bundle and let $R$ and $R'$ be the distance between the two muons and the bundle axis. $R'$ can be obtained from the distance $D$ and the angle $\phi$ between the two muons
\beq
R'(R,D,\phi)=\sqrt{R^2+D^2-2RD\textrm{cos}(\phi)} .
\eeq
In order to calculate $f'(D)$, we consider the distribution of the muons per element of area normal to the axis
\beq
\frac{dN}{dA}=\frac{f(R)}{2\pi R}.
\eeq
The distribution $f'(D)$ is then the integral
\beq
f'(D)=\int_{R=0}^{\infty}\int_{\phi=0}^{2\pi} f(R) f(R'(R,D,\phi))\frac{D}{2\pi R'(R,D,\phi)} \textrm{d}R \textrm{d}\phi .
\eeq
The probability that the detector is hit by both muons in a 2-muon bundle is
\begin{eqnarray}
&&P_{M\mu}=\frac{1}{R(2)}\int_{b=0}^{R_d}\textrm{d}b\int_{\theta=0}^{\pi/2}4\pi^2 b\Phi(h,\theta,2)\textrm{tan}(\theta) \textrm{d}\theta \nonumber\\ &&\left(\int_{D=0}^{R_d-b}f'(D) \textrm{d}D + \int_{D=R_d-b}^{R_d+b}f'(D)\textrm{arccos}\left(\frac{b^2 + d^2 - R_d^2}{2bD}\right) \textrm{d}D \right)
\end{eqnarray}
where $R_d$ is the radius of the detector.  We evaluate this quantity numerically for each candidate site.

In Table~\ref{ratetab} we summarize $P_{M\mu}$ together with several other relevant quantities.  The muon rate is defined to be $\sum_{m=1}^{\infty} mR(m)$.  The quantity $\sum_{m=1}^{\infty} R(m)$ is the sum of the single muon and bundle rates.  This systematically underestimates the event rate by about 3-5\% because, as we have seen, sometimes only some of the muons from a given bundle strike the detector.   However this correction is neglible compared to our uncertainties, which are around 20\%, and so we will simply refer to this quantity as the event rate.  The bundle rate is $\sum_{m=2}^{\infty} R(m)$. The $m$-muon rate is just $R(m)$ while the mean muon energy is calculated for single muons with energy below about 5 TeV.  The latter is higher than typical values given in the literature as a result of the long high energy tail which consists of few muons but leads to a significant fraction of the isotope production.  

\begin{figure} %[!tph]
\begin{center}
\includegraphics[width=4.2in]{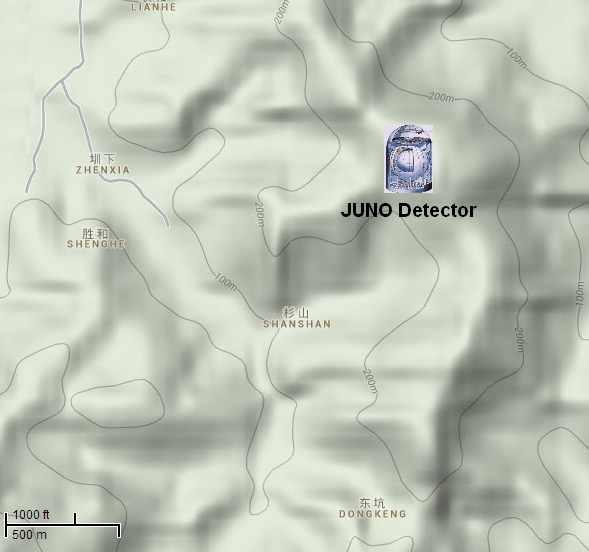}
\includegraphics[width=4.2in]{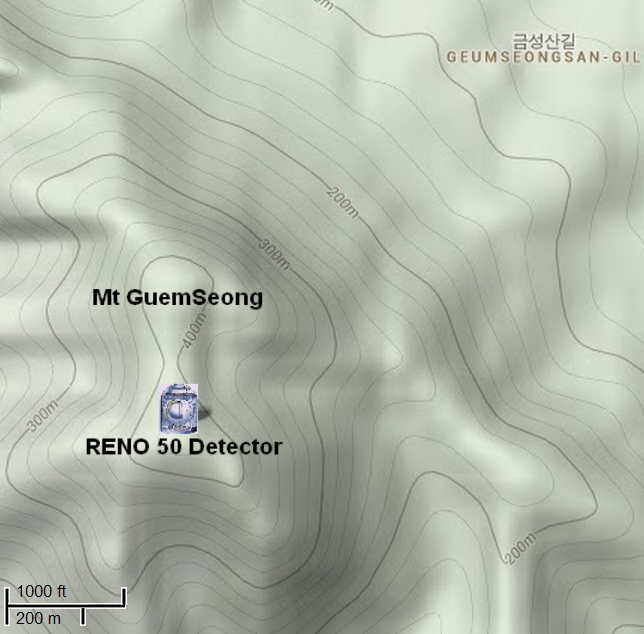}
\caption{Google maps illustrating the favored locations of the JUNO and RENO 50 detectors}
\label{mappe}
\end{center}
\end{figure}

We have considered a detector 700 and 900 meters underneath a flat surface as well as various topographies corresponding to potential sites for detectors.  In particular we have considered the preferred Dong Keng \cite{noisim} site for JUNO and the preferred Mt GuemSeong \cite{noisim} site for RENO 50, which are illustrated in the upper and lower panels of Fig.~\ref{mappe} respectively.  These are respectively $h_r=700$ meters and $900$ meters underneath the peaks of their corresponding hills, where $h_r$ is the rock overburden which is equal to $h_r=h/2.64$.  We have also considered a location 200 meters beneath the preferred site for JUNO.   Note that the reported mean muon energies are higher than that found in other studies.  This increase is caused by a small number of muons in the very high energy tail of the cosmogenic muon distribution, at several TeV.  

To determine the expected muon flux ideally one requires a geological survey of the rocks around these sites and a full 3d simulation such as MUSIC.   We have instead simply assumed that the rock is standard and have employed a cylindrically-symmetric approximation to the topographies illustrated in Fig.~\ref{mappe}.  For JUNO we have assumed an overburden of $h_r=700$ meters of rock in a 100 meter radius circle, followed by an annulus of inner radius 100 meters and outer radius 500 meters with a surface 50 meters lower, than another annulus of outer radius 700 meters with a surface 50 meters yet lower and finally we have assumed that the surface is another 50 meters lower at radii beyond 700 meters.  Similarly we have approximated Mt Guemseong using a series of annuli, beginning with a 200 meter radius circle with $h_r=900$ meters of rock overburden, followed by a 600 meter annulus with $h_r=800$ meters of overburden, a 900 meter radius annulus with $h_r=700$ meters of overburden and then we have assumed a $h_r=600$ meter overburden at all radii beyond 900 meters.  %Note that the values of $h$ used in Eqs.~(\ref{keq,phieq}) are equal to the rock overburdens listed in this paragraph multiplied by 

\begin{table}[position specifier]
\centering
\begin{tabular}{c|l|l|l|l|l}
&700 m&900 m&JUNO&JUNO+200m&RENO 50\\
\hline\hline
%40 km&&\%\ (\%)&\%\ (\%)\\
%\hline
muon rate&$3.0\pm 0.7$&$1.1\pm 0.2$&$5.4\pm 1.2$&$1.9\pm 0.4$&$3.1\pm 0.6$\\
\hline
event rate&$2.3\pm 0.5$&$0.90\pm 0.19$&$4.1\pm 0.9$&$1.5\pm 0.3$&$2.4\pm 0.5$\\
\hline
bundle rate&$0.36\pm 0.09$&$0.12\pm 0.03$&$0.69\pm 0.16$&$0.22\pm 0.05$&$0.37\pm 0.08$\\
\hline
single $\mu$ rate&$2.0\pm 0.4$&$0.78\pm 0.16$&$3.4\pm 0.7$&$1.3\pm 0.3$&$2.1\pm 0.4$\\
\hline
two $\mu$ rate&$0.23\pm 0.06$&$0.081\pm 0.019$&$0.43\pm 0.10$&$0.14\pm 0.03$&$0.24\pm 0.05$\\
\hline
$P_{M\mu}$& $0.45$ & $0.54$ & $0.40$ & $0.49$ &$0.45$\\
\hline
three $\mu$ rate&$0.067\pm 0.017$&$0.022\pm 0.005$&$0.13\pm 0.03$&$0.040\pm 0.010$&$0.069\pm 0.016$\\
\hline
mean $\mu$ energy&$267\pm 8$&$310\pm 8$&$254\pm 7$&$294\pm 7$&$284\pm 7$\\
\end{tabular}
\caption{Rates in Hz of various kinds of events under 700/900 meters of rock with a flat surface, at JUNO, 200 meters beneath JUNO and at RENO 50 and the mean energy in GeV.  The errors do not include systematic errors in the neutrino model, but consist of an uncertainty of an overall 30 meters in the surface elevation and a 5\% uncertainty in the rock density.  The $n$ muon rate is the rate of muons striking the detector which are in $n$-muon bundles, regardless of how many of the other $n-1$ muons actually enter the detector.  The quantity $P_{M\mu}$ is the probability that both muons in a 2-muon bundle hit the detector.}
\label{ratetab}
\end{table}

In Table~\ref{ratetab} one can see that at JUNO one expects that for 40\% of all 2-muon events in which a single muon strikes detector, both muons will strike the detector.  As 2-muon events occur at a rate of 0.43 Hz, this means that they contribute 0.17 Hz to the bundle rate.  Events with more than 2 muons occur at a rate of 0.26 Hz.  If we make the crude approximation that in the case of all such events, at least 2 muons enter the detector then we find a multi-muon event rate of 0.43 Hz which is 10\% of the total event rate of 4.4 Hz, where we have added 0.3 Hz corresponding to the fact that 60\% of 2-muon bundles appear as two events, with one muon from each of two bundles.  However although only 10\% of events are multimuon events, these account for 1.5 Hz of the 5.4 Hz total muon rate and so 28\% of all muons.   Thus one expects about 28\% of all cosmogenic muon isotope backgrounds to result from multimuon events, where localized vetoes require very difficult tracking.

\section{Isotope Production} \label{issez}

\begin{figure} %[!tph]
\begin{center}
\includegraphics[width=4.0in]{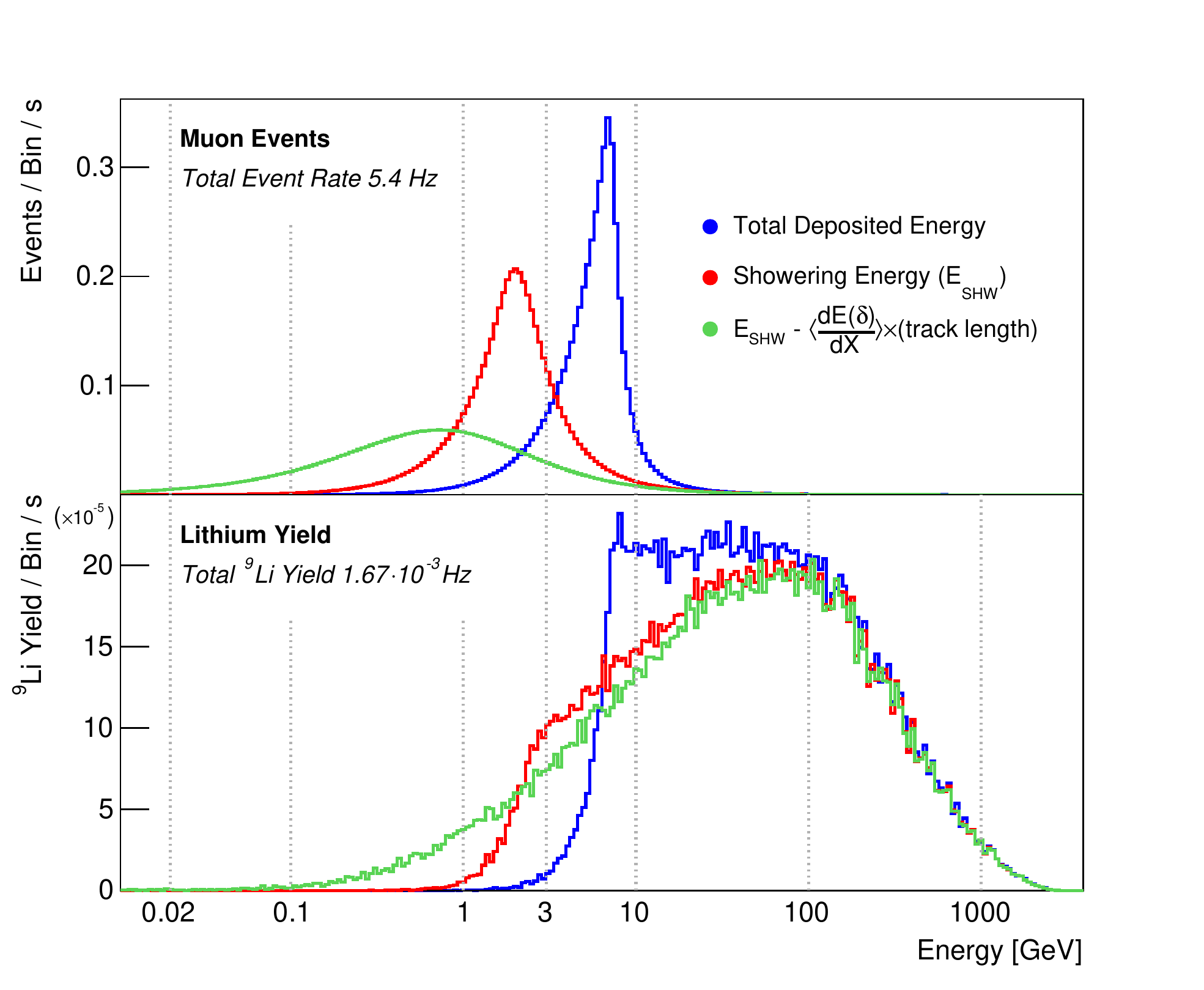}
\caption{Top: The spectrum of the deposited total energy, showering and showering minus expected $\delta$ ray energy at JUNO.  Bottom:  The rate of ${}^9$Li production as a function of the total deposited, showering energy and $\delta$ ray subtracted showering energies.  Note that most ${}^9$Li is produced by the very high energy tail of the energy deposition (above 10 GeV) and so is easily cut.}
\label{lispectra}
\end{center}
\end{figure}

\begin{figure} %[!tph]
\begin{center}
\includegraphics[width=6in]{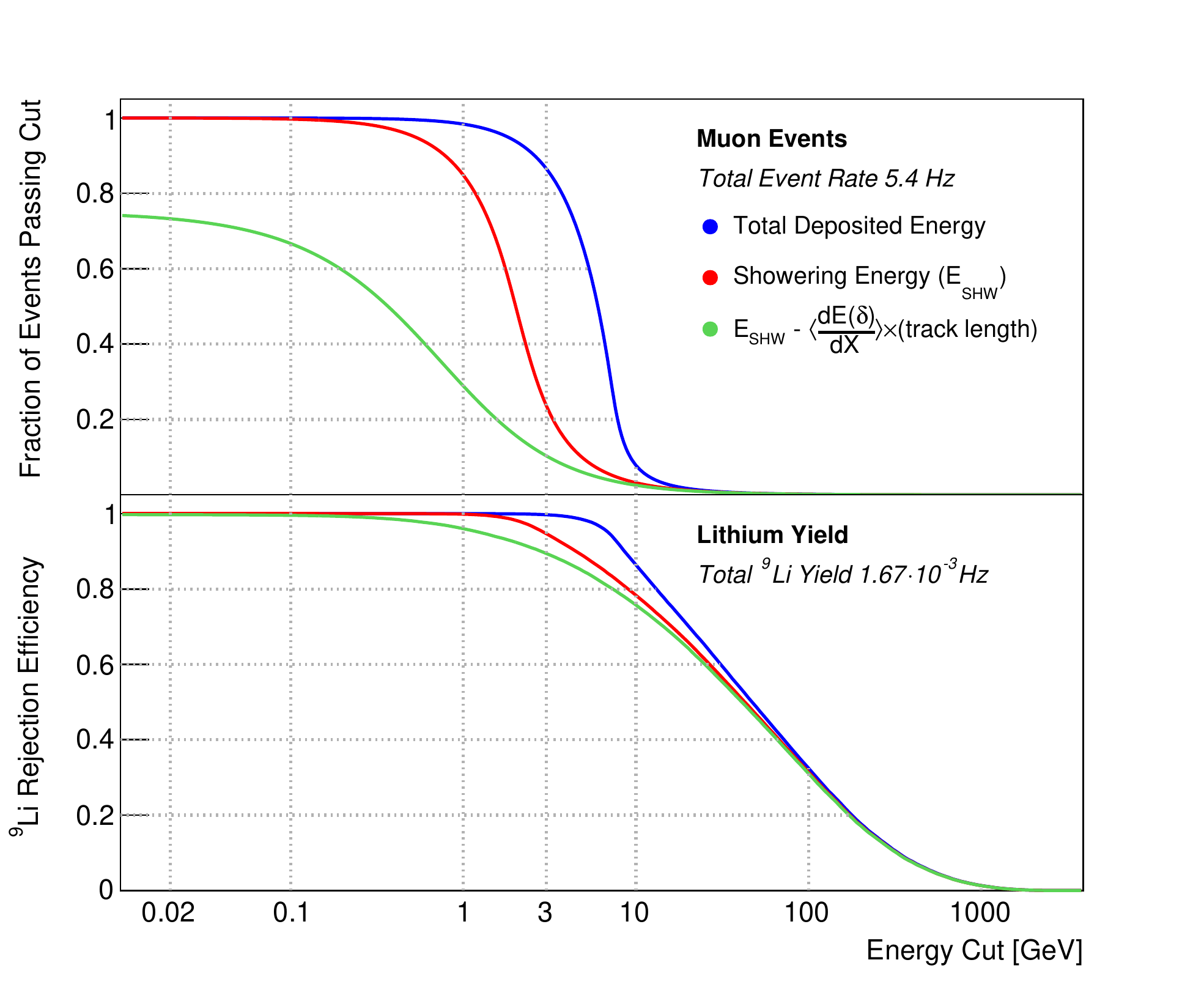}
\caption{Top: The fraction of muons for which the deposited energy exceeds the value in the horizontal axis.  Bottom: The fraction of the total ${}^9$Li production resulting from muons which deposit more energy than the threshold reported on the horizontal axis.  For example, 78\% of the total ${}^9$Li production results from those muons with a showering energy greater than 10 GeV.  This number provides an upper bound on the rejection efficiency that may be obtained by vetoing events occurring after the passage of such muons. The blue curves are defined by identifying the horizontal axis with the total deposited energy, the red curves use the showering energy while the green curves use the showering energy minus the expected $\delta$ ray deposition given a fixed track length.}
\label{cumfig}
\end{center}
\end{figure}

In Fig. \ref{lispectra} we present the spectra of deposited total energy and deposited showering energy expected at JUNO.  We also plot the deposited showering energy minus the most likely deposited $\delta$ ray energy, which is $0.6$ MeV/cm multiplied by the track length.  In the lower panel we show the ${}^9$Li rate as a function of deposited total energy, showering energy and $\delta$ ray subtracted total energy.    In Fig.~\ref{cumfig} we plot, in the top panel, the fraction of the muons whose energy deposition exceeds a given threshold and in the bottom panel one sees the fraction of ${}^9$Li produced by these muons.  This plot is obtained by integrating and renormalizing Fig.~\ref{lispectra}.

One can again see that a 3 GeV cut on the showering energy would require a veto after about 23\% of all muon events, which if longer than the ${}^9$Li half life would lead to a large dead time.  It would however reject about 95\% of ${}^9$Li production.  On the other hand, a cut near 10 GeV would require a veto of only 3\% of events but only reject 78\% of the ${}^9$Li.  A veto based on the total energy deposition appears mildly more problematic, a 95\% rejection efficiency requires a veto of the 35\% of all muons with total energy deposition above 6.9 GeV.  If the most common $\delta$ ray energy at the corresponding track length is subtracted from the showering energy, one obtains a 95\% rejection efficiency at the price of vetoing the 24\% of all muon events with a $\delta$ ray subtracted showering energy beyond 1.2 GeV.   This leads to a dead time with fixed rejection efficiency which agrees, to within the precision introduced by our binning, with that obtained using a showering energy cut and no $\delta$ ray subtraction.

\begin{figure} %[!tph]
\begin{center}
\includegraphics[width=4.0in]{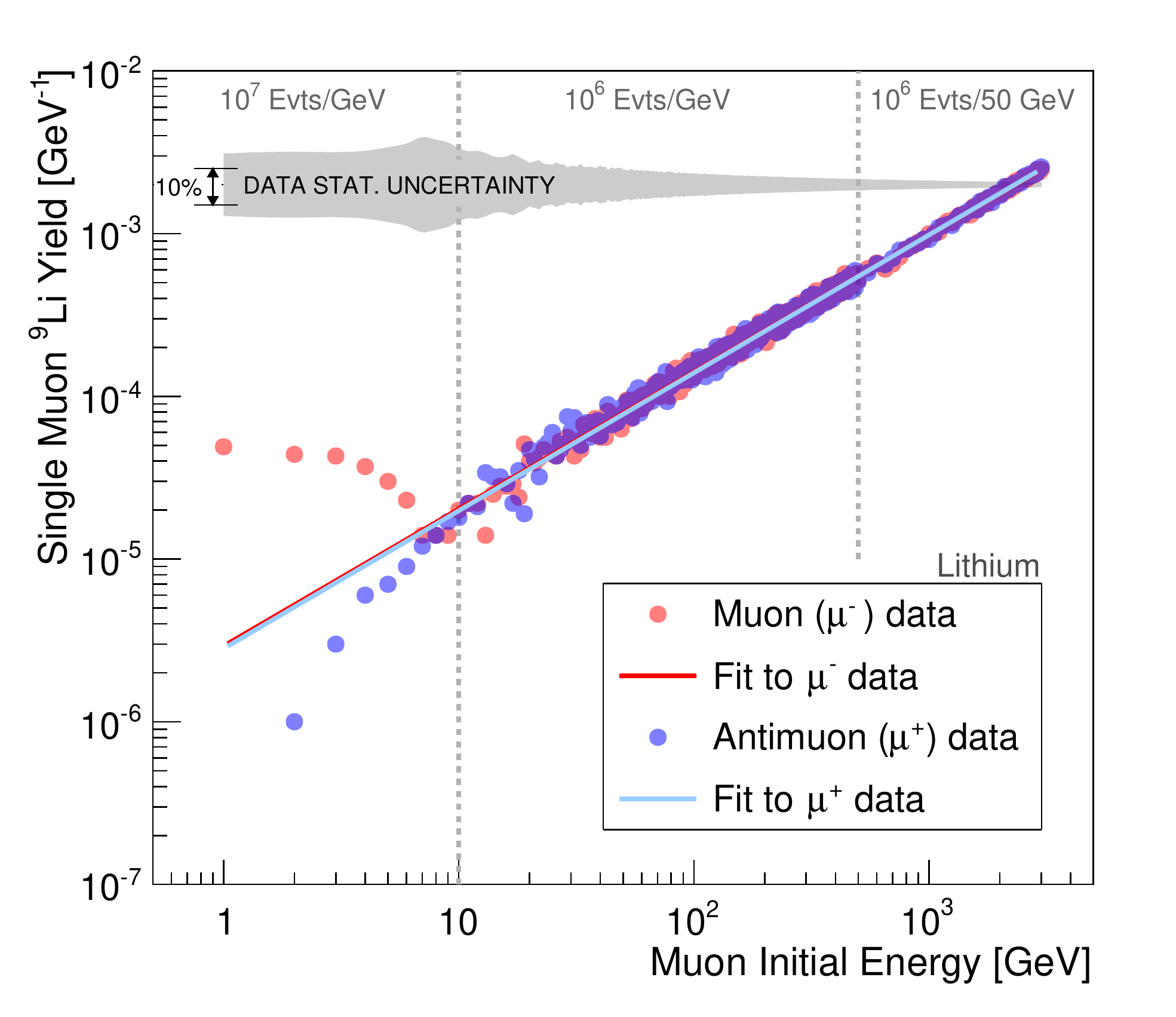}
\includegraphics[width=4.0in]{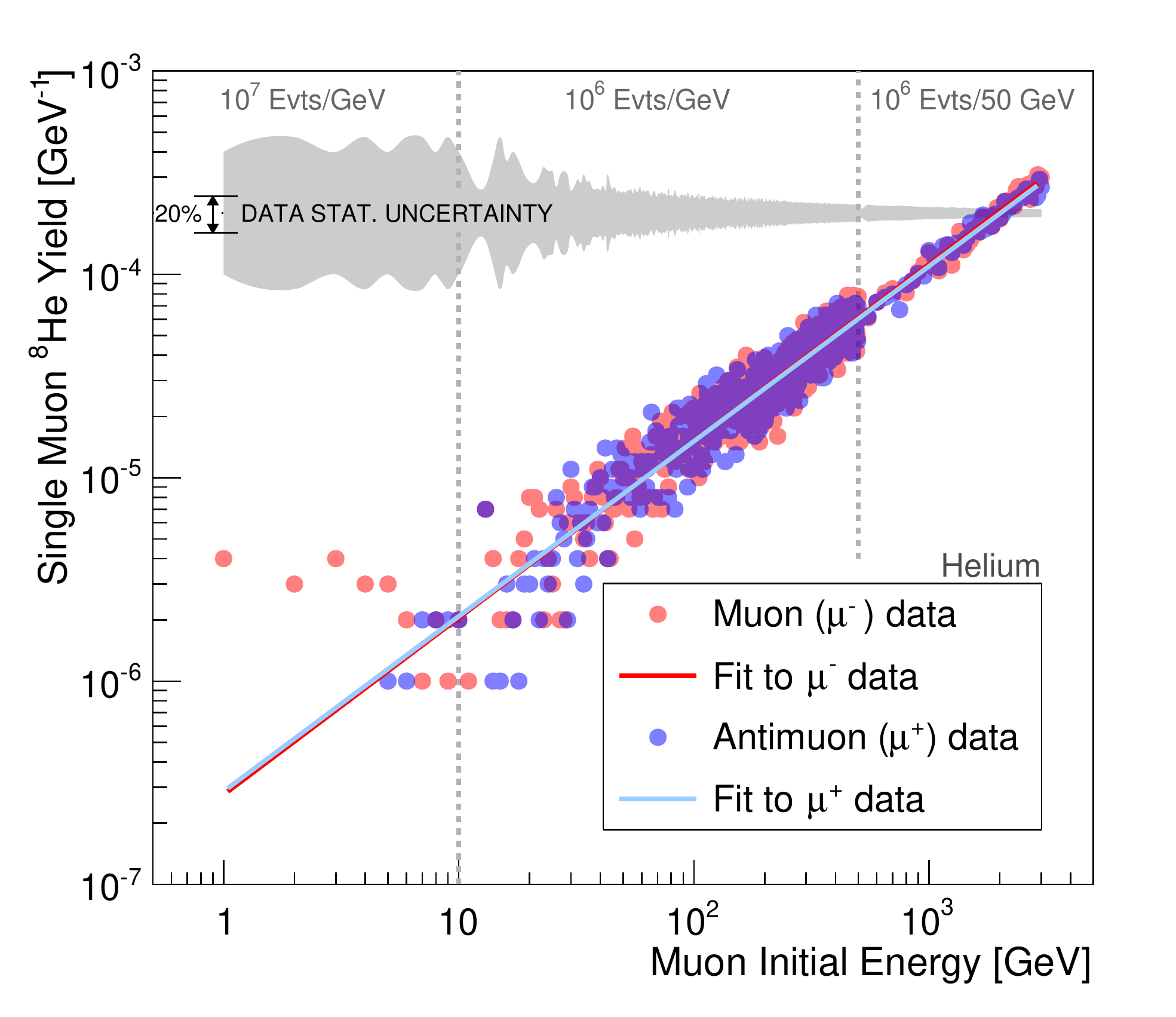}
\caption{The mean number of ${}^9$Li (top) and ${}^8$He (bottom) created by a single $\mu^-$ (red) and $\mu^+$ (blue) with a given energy.  The lines are the best power law fits. The gray band represents the relative statistical uncertainty on the simulated isotope yield.}
\label{liperenergia}
\end{center}
\end{figure}

We have also estimated the total yield of ${}^9$Li and ${}^8$He at each experimental site.  To do this, we ran $10^7$ FLUKA simulations of single monoenergetic muons  of each energy between 1 GeV and 10 GeV and $10^6$ simulations at energy between 11 GeV and 500 with a step size of 1 GeV, with an additional $10^6$ simulations at each energy up to 3 TeV with a step size of 50 GeV, for a total of $6.4 \times 10^8$ simulations.  Again the impact parameter was randomized with a distribution reflecting a homogeneous distribution of muons.  The   ${}^9$Li yield and best power law fit, with an exponent of $0.842\pm 0.002$ are presented in red in the top panel of Fig.~\ref{liperenergia}.  The results of another $6.4\times 10^8$ simulations of antimuons are presented in blue, together with a power law fit whose exponent is $0.847\pm 0.002$.  The same simulations also determined the ${}^8$He rate, as is shown on the bottom panel.  The best fit exponents are $0.869\pm 0.006$ and $0.861\pm 0.006$ for $\mu^-$ and $\mu^+$ respectively.  

The power law fits reproduce the simulated isotope yields to within the statistical errors from 10 GeV up to 3 TeV, where the vast majority of the spallation isotope production occurs.  However below 10 GeV, the isotope production from $\mu^-$ events increases.  Indeed in this range the ${}^9$Li yield is 3.3 times greater for $\mu^-$ events than for $\mu^+$.  This is because $\mu^-$ at these low energies stop and bind with the hydrogen and carbon in the scintillator.  The deeper potential well about the carbon nucleii means that any $\mu^-$ which initially bind to hydrogen are soon transfered to carbon and these quickly decay to 1s or 1p orbitals.  Here most of the $\mu^-$, like the $\mu^+$, simply decay.  However 7\% of the $\mu^-$ are then captured by the carbon nucleii, undergoing a charged current interaction which can create spallation isotopes.

\begin{figure} %[!tph]
\begin{center}
\includegraphics[width=4.0in]{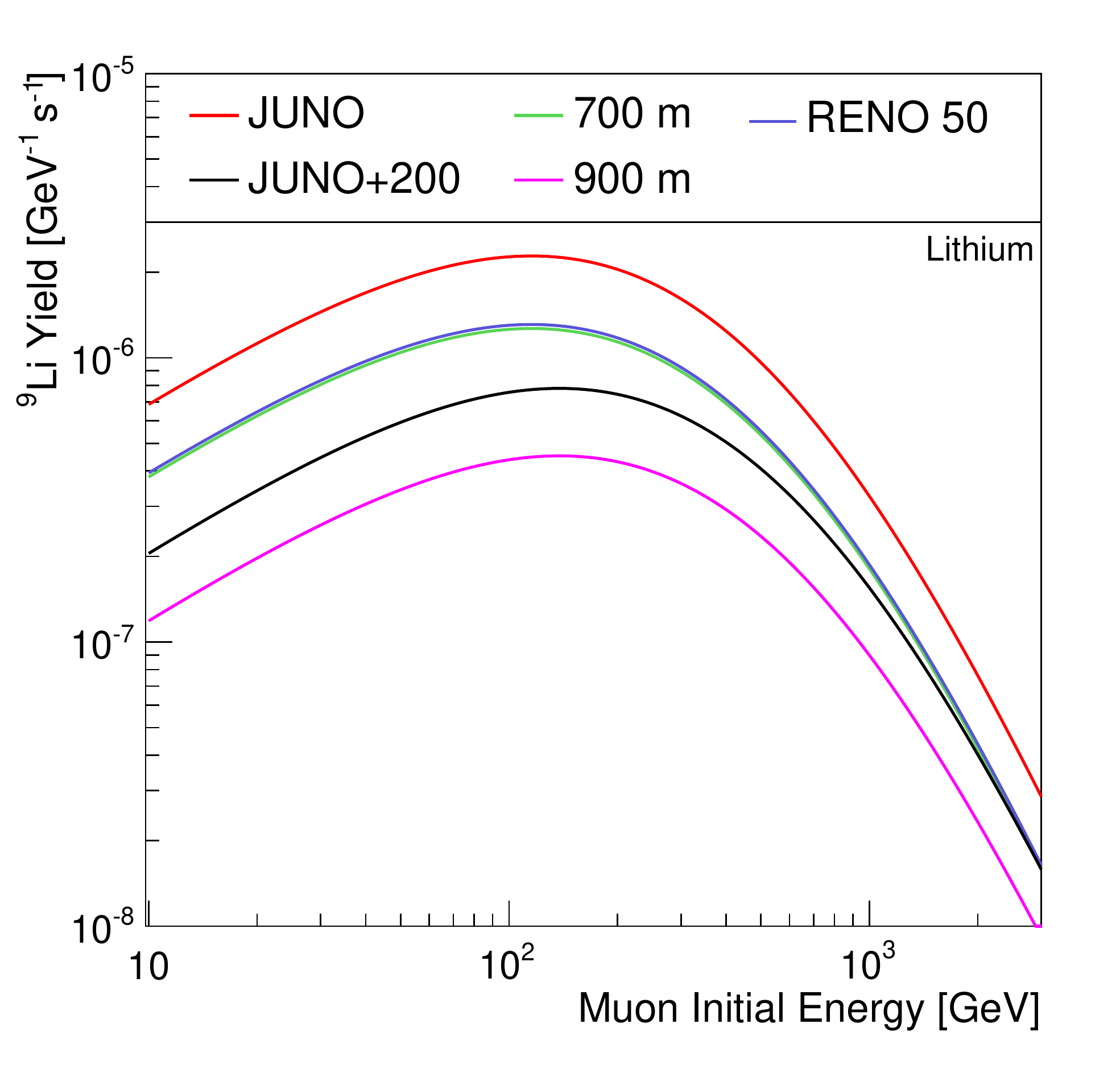}
\includegraphics[width=4.0in]{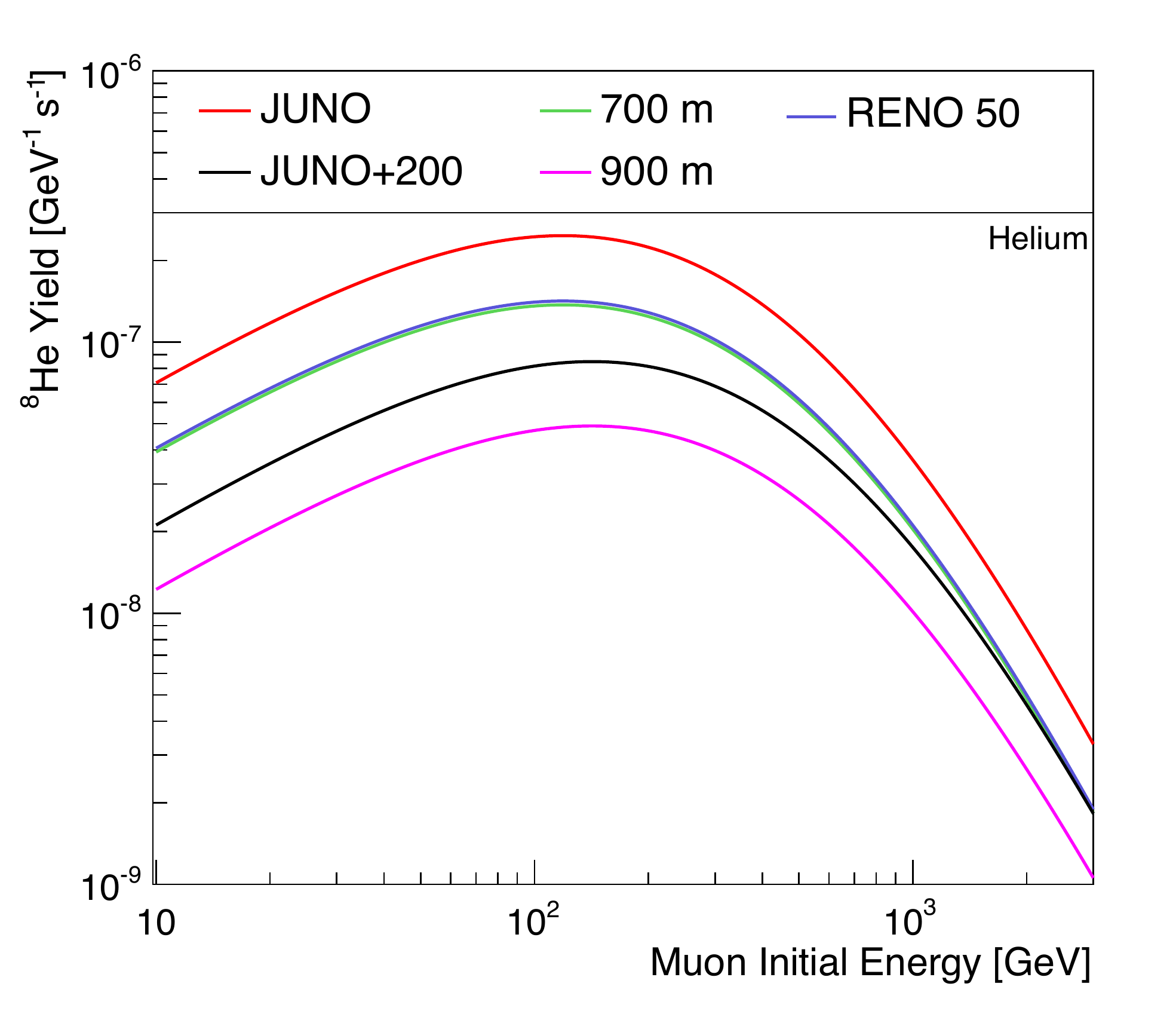}
\caption{The ${}^9$Li (top) and ${}^8$He (bottom) production rates per energy of the cosmogenic muon at the five experimental sites}
\label{li}
\end{center}
\end{figure}

To arrive at the rate for each experimental site we have folded the power law fits to the simulated data as a function of muon energy with the expected muon spectra reported in Fig.~\ref{initfig}.  In contrast with Fig.~\ref{initfig}, here we are interested in the total $\mu$ rate, not just those arising from single $\mu$ events, so we have rescaled this result by the ratio of the total $\mu$ rate to the single $\mu$ rate reported in Table~\ref{ratetab}.  We have assumed that the ratio of $\mu^+$ to $\mu^-$ is $1.37$, as was found for 1.2 TeV muons by Kamiokande in Ref.~\cite{antimu}.  The resulting ${}^9$Li and ${}^8$He rates per muon energy bin are displayed in Fig.~\ref{li}.  Next, to arrive at the total ${}^9$Li and ${}^8$He rates we have integrated this figure over the $\mu$ energy.  The resulting total rates at each experimental site are summarized in Table~\ref{litab}.  The  ${}^9$Li  rates are compatible with the rough estimate of Ref.~\cite{menufact} obtained by simply rescaling KamLAND's rate.

As can be seen in Table~\ref{litab}, the total spallation isotope rates are appreciably higher than the expected reactor neutrino IBD signal rates of $3\times 10^{-4}$ Hz for JUNO and $2\times 10^{-4}$ Hz for RENO 50 at Guemseong.  However only those decays which produce a neutron yield a false double coincidence signal.  These are $51\%$ of ${}^9$Li decays and only 16\% of ${}^8$He decays.  In the bottom two rows of Table~\ref{litab} we report the false double coincidence rate expected.  Note that this background double coincidence rate is still three times greater than the IBD signal rate at JUNO and RENO 50, although it would be only slightly greater than the RENO 50 signal rate at the Munmyeong site of Ref.~\cite{2rivelsim} even with the same overhead burden as the Guemseong site.

In practice the background rate will be reduced, at the expense of dead time, by a veto program.  Similarly, the expected background can be subtracted from a signal by a shape analysis.  However the statistical fluctuations created by the background will survive this subtraction and obscure the low energy oscillations in the reactor neutrino spectrum whose observation is necessary for a determination of the hierarchy.  Thus, both the sensitivity to the hierarchy and the precision with which $\theta_{12}$ may be measured depend critically on the optimal choice of muon vetoes.  As the muon rate is comparable to the ${}^9$Li  half life, full detector vetoes are not an option for all but the highest energy showers, thus excellent muon tracking, including the tracking of multimuon events and even muons that arrive horizontally will be necessary to achieve these science goals.  

\begin{table}[position specifier]
\centering
\begin{tabular}{l|c|c|c|c|c}
                    &       700 m           &  900 m            &  JUNO             &  JUNO+200m     & RENO 50\\
\hline\hline
${}^9$Li  rate      &       $93\pm20        $&$ 39.3\pm 8.2     $&$ 167\pm 37       $&$ 68\pm 14     $&$ 96\pm 20$\\
\hline
${}^8$He rate       &       $10.3\pm 2.1    $&$ 4.37\pm 0.92    $&$ 18.5\pm 4.1     $&$ 7.5\pm 1.5   $&$ 10.5\pm 2.1$\\
\hline
${}^9$Li  $n$-decay rate&   $47\pm 10       $&$ 20.0\pm 4.2     $&$ 85\pm 19        $&$ 34.6\pm 7.3  $&$ 49\pm 10$\\
\hline
${}^8$He $n$-decay rate&    $1.65\pm 0.35   $&$ 0.70\pm 0.15    $&$ 2.96\pm 0.65    $&$ 1.20\pm 0.24 $&$ 1.69\pm 0.35$\\
\hline
\end{tabular}
\caption{Rates in $10^{-5}$Hz of ${}^9$Li and ${}^8$He production via cosmogenic muon spallation on ${}^{12}$C under 700/900 meters of rock with a flat surface, at JUNO, 200 meters beneath JUNO and at RENO 50.  The resulting ${}^9$Li and ${}^8$He decays only provide false double coincidence signals when their decay produces a neutron, which happens in $51\%$ and $16\%$ of their decays respectively.  The last two rows report the rates of these decays.   The errors reported reflect only the uncertainty in the muon rate, not the uncertainties in the isotope production per muon.}
\label{litab}
\end{table}

\section{Concluding Remarks}

As can be seen in the lower panel of Fig.~\ref{finalfig}, for any depth in the range relevant to future large liquid scintillator experiments, about 20\% of single muons will be showering in a 20 kton spherical detector.  Thus to determine the showering single muon rate, one only needs to find the single muon rate for a given site and multiply it by 20\%.  Similarly from Table~\ref{ratetab} we can see that 17-20\% of muon events will occur in muon bundles.  As, for the large detectors considered here, in most cases muon bundles yield multimuon events in the detector, the multimuon event rate is about 10\%.   Again, this bundle fraction is fairly robust against changes in depth and topography and so may be applied to a wide variety of candidate experimental sites.

One immediate consequence of our study is that both the bundle rates are roughly 0.8 Hz (0.5 Hz) at JUNO (RENO 50) with the shower rates only about 40\% lower.  This means that KamLAND's veto strategy, employing 2 second full detector cuts for showering and poorly constructed tracks, cannot be applied to these experiments as the events are separated by less than 2 seconds.  If JUNO is moved 200 meters lower, KamLAND's cuts would still be problematic although 1 second full detector cuts would be possible with considerable dead time.  1 second cuts are sufficient to reduce the ${}^9$Li and ${}^8$He backgrounds well below the level of the signal.  KamLAND's veto strategy has been assumed in all studies of these backgrounds that have so far appeared in the literature, and so our results indicate that these studies need to be redone.

As a result, full detector cuts will be infeasible at these experiments and it is critical that muon tracking be successful for the vast majority of single and multimuon events, so that selective cuts may remove events from cylinders surrounding cosmogenic muon tracks.  A muon veto system above the main detector will help with downgoing muons.  However, a flat 1300 m${}^2$ detector placed 5 meters above the JUNO (RENO 50) inner detector will only be exposed to 40\% (37\%) of the muons that eventually pass through the inner detector and these will in general be somewhat less energetic and so produce less isotopes than average.  

In our next paper we will investigate the creation of ${}^9$Li and ${}^8$He and determine the effectiveness of various veto schemes  as well as the cost to the scientific goals both as a result of the increased dead time and as a result of the background which will mask the 1-3 oscillations in the 3-4 MeV range whose detection is essential to determine the neutrino mass hierarchy \cite{noiteor}.  However for any proposed veto scheme, the lower panel of Fig.~\ref{finalfig} together with the total muon flux may be used to calculate the resulting veto efficiency and so the resulting dead time for the experiment.  

\section* {Acknowledgement}
\noindent
We are pleased to thank Xin Qian and Anton Empl for suggestions and correspondence.  JE is supported by NSFC grant 11375201 and a KEK fellowship.  EC  is supported by the Chinese Academy of Sciences Fellowship for Young International Scientists grant  number 2013Y1JB0001 and NSFC grant 11350110500.  XZ is supported in part by  NSFC grants 11121092, 11033005 and 11375202.   

%%%%%%%%%%%%%%%%%%%%%%%%%%%%%%%%

\end{document}

\bibitem{2rivel}
   E.~Ciuffoli, J.~Evslin and X.~Zhang,
  %``Mass Hierarchy Determination Using Neutrinos from Multiple Reactors,''
  JHEP {\bf 1212} (2012) 004
  [arXiv:1209.2227 [hep-ph]].
  E.~Ciuffoli, J.~Evslin, Z.~Wang, C.~Yang, X.~Zhang and W.~Zhong,
  %``Medium Baseline Reactor Neutrino Experiments with 2 Identical Detectors,''
  arXiv:1211.6818 [hep-ph] and
  %``Advantages of Multiple Detectors for the Neutrino Mass Hierarchy Determination at Reactor Experiments,''
  arXiv:1308.0591 [hep-ph].

\bibitem{daed}
  J.~Alonso, F.~T.~Avignone, W.~A.~Barletta, R.~Barlow, H.~T.~Baumgartner, A.~Bernstein, E.~Blucher and L.~Bugel {\it et al.},
  %``Expression of Interest for a Novel Search for CP Violation in the Neutrino Sector: DAEdALUS,''
  arXiv:1006.0260 [physics.ins-det].

\bibitem{lena}
 M.~Wurm {\it et al.}  [LENA Collaboration],
  %``The next-generation liquid-scintillator neutrino observatory LENA,''
  Astropart.\ Phys.\  {\bf 35} (2012) 685
  [arXiv:1104.5620 [astro-ph.IM]].

\bibitem{whitepaper}
  C.~Aberle, A.~Adelmann, J.~Alonso, W.~A.~Barletta, R.~Barlow, L.~Bartoszek, A.~Bungau and A.~Calanna {\it et al.},
  %``Whitepaper on the DAEdALUS Program,''
  arXiv:1307.2949 [physics.acc-ph].

\bibitem{lsnd}
  A.~Aguilar-Arevalo {\it et al.}  [LSND Collaboration],
  %``Evidence for neutrino oscillations from the observation of anti-neutrino(electron) appearance in a anti-neutrino(muon) beam,''
  Phys.\ Rev.\ D {\bf 64} (2001) 112007
  [hep-ex/0104049].

\end{thebibliography}


\begin{thebibliography}{99}%\setlength{\itemsep}{-2.3mm}

%%%%%%%%%%%%%%%%%%%%%%%%%%%%%%%%%
\bibitem{juno}
  Y.-F.~Li, J.~Cao, Y.~Wang and L.~Zhan,
  ``Unambiguous Determination of the Neutrino Mass Hierarchy Using Reactor Neutrinos,''
  Phys.\ Rev.\ D {\bf 88} (2013) 013008
  [arXiv:1303.6733 [hep-ex]].

\bibitem{reno50}
 S.~B.~Kim, 
``Proposal for RENO-50; detector design \& goals,"
talk given at the {\it International Workshop on RENO-50: Towards the Neutrino Mass Hierarchy} at Seoul National University.

\bibitem{lena}
 M.~Wurm {\it et al.}  [LENA Collaboration],
  ``The next-generation liquid-scintillator neutrino observatory LENA,''
  Astropart.\ Phys.\  {\bf 35} (2012) 685
  [arXiv:1104.5620 [astro-ph.IM]].
 
\bibitem{cr}
 F.~Reines and C.~L.~Cowan,
  ``A Proposed experiment to detect the free neutrino,''
  Phys.\ Rev.\  {\bf 90} (1953) 492.

\bibitem{kamlandback}
  S.~Abe {\it et al.}  [KamLAND Collaboration],
  ``Production of Radioactive Isotopes through Cosmic Muon Spallation in KamLAND,''
  Phys.\ Rev.\ C {\bf 81} (2010) 025807
  [arXiv:0907.0066 [hep-ex]].

\bibitem{fluka}
G.~Battistoni, S.~Muraro, P.~R.~Sala, F.~Cerutti, A.~Ferrari,
S.~Roesler, A.~Fass\`o, J.~Ranft,
Proceedings of the Hadronic Shower Simulation Workshop 2006,
Fermilab 6-8 September 2006, M.~Albrow, R.~Raja eds., 
AIP Conf Proceedings {\bf 896} (2007) 31. %-49.
``FLUKA: a multi-particle transport code"
A.~Ferrari, P.~R. Sala, A.~Fass\`o, and J.~Ranft,
CERN-2005-10 (2005), INFN/TC\_05/11, SLAC-R-773 

\bibitem{bundle}
  Y.~Becherini, A.~Margiotta, M.~Sioli and M.~Spurio,
  ``A Parameterisation of single and multiple muons in the deep water or ice,''
  Astropart.\ Phys.\  {\bf 25} (2006) 1
  [hep-ph/0507228].

\bibitem{lindley}
  L.~A.~Winslow,
  ``First Solar Neutrinos from KamLAND: A Measurement of the B-8 Solar Neutrino Flux,''
  ISBN-9781109098198.

\bibitem{antimu}
 M.~Yamada, K.~Miyano, H.~Miyata, H.~Takei, M.~Mori, Y.~Oyama, A.~Suzuki and K.~Takahashi {\it et al.},
  ``Measurements of the charge ratio and polarization of 1.2-TeV/c cosmic ray muons with the KAMIOKANDE-II detector,''
  Phys.\ Rev.\ D {\bf 44} (1991) 617.

\bibitem{dwyer}
  D. A. Dwyer,
  ``Precision Measurement of Neutrino Oscillation Parameters with
KamLAND,''
  PhD Thesis, UC Berkeley, Spring 2007.


\bibitem{menufact}
  J.~Evslin,
  ``Confidence in the neutrino mass hierarchy,''
  arXiv:1310.4007 [hep-ph].

\bibitem{noisim}
  E.~Ciuffoli, J.~Evslin and X.~Zhang,
  ``Mass Hierarchy Determination Using Neutrinos from Multiple Reactors,''
  JHEP {\bf 1212} (2012) 004
  [arXiv:1209.2227 [hep-ph]].

\bibitem{noiteor}
 E.~Ciuffoli, J.~Evslin and X.~Zhang,
  ``The Neutrino Mass Hierarchy at Reactor Experiments now that $\theta_{13}$ is Large,''
  JHEP {\bf 1303} (2013) 016
  [arXiv:1208.1991 [hep-ex]].

\bibitem{2rivelsim}
E.~Ciuffoli, J.~Evslin, Z.~Wang, C.~Yang, X.~Zhang and W.~Zhong,
  ``Advantages of Multiple Detectors for the Neutrino Mass Hierarchy Determination at Reactor Experiments,''
  arXiv:1308.0591 [hep-ph].

\end{thebibliography}
\end{document}